\documentclass[10pt,preprint]{aastex}
\usepackage{natbib}
\usepackage{multirow}
\usepackage{upgreek}
\usepackage[colorlinks=true, pdfstartview=FitV, linkcolor=blue,citecolor=blue, urlcolor=blue]{hyperref}
\usepackage{wasysym}
\usepackage{txfonts}
\usepackage{gensymb}
\usepackage{morefloats}
\usepackage{threeparttable}
\slugcomment{version \today}
\shorttitle{Hidden AGNs in Early-Type Galaxies}
\shortauthors{Paggi et al.}

\begin{document}

\title{Hidden AGNs in Early-Type Galaxies}

\author{Alessandro Paggi\altaffilmark{1}, Giuseppina 
Fabbiano\altaffilmark{1}, Francesca Civano\altaffilmark{1,2}, Silvia 
Pellegrini\altaffilmark{3}, Martin 
Elvis\altaffilmark{1} and Dong-Woo Kim\altaffilmark{1}}
\affil{\altaffilmark{1}Harvard-Smithsonian Center for Astrophysics, 60 
Garden St, Cambridge, MA 02138, USA: 
\href{mailto:apaggi@cfa.harvard.edu}{apaggi@cfa.harvard.edu}\\
\altaffilmark{2}Department of Physics and Yale Center for Astronomy and 
Astrophysics, Yale University, P.O. Box 208121, New Haven, CT 06520-8121\\
\altaffilmark{3}Department of Astronomy, University of Bologna, via Ranzani 
1, 40127 Bologna, Italy}

\begin{abstract}
We present a stacking analysis of the complete sample of Early Type Galaxies 
(ETGs) in the \textit{Chandra} COSMOS (C-COSMOS) survey, to explore the nature 
of the X-ray luminosity in the redshift and stellar luminosity ranges 
\(0<z<1.5\) and \({10}^{9}<L_K/L_{\astrosun}<{10}^{13}\). Using established 
scaling relations, we subtract the contribution of X-ray binary populations, to 
estimate the combined emission of hot ISM and AGN. To discriminate between the 
relative importance of these two components, we (1) compare our results with the 
relation observed in the local universe \(L_{X,gas}\propto L_K^{4.5}\) for hot 
gaseous halos emission in ETGs, and (2) evaluate the spectral signature of each 
stacked bin. We find two regimes where the non-stellar X-ray emission is hard, 
consistent with AGN emission. First, there is evidence of hard, absorbed X-ray 
emission in stacked bins including relatively high z (\(\sim 1.2\)) ETGs with 
average high X-ray luminosity (\(L_{X-LMXB}\gtrsim 6\times{10}^{42}\mbox{ 
erg}/\mbox{s}\)). These luminosities are consistent with the presence of
highly absorbed ``hidden" AGNs in these ETGs, which are not visible in their 
optical-IR spectra and spectral energy distributions. Second, confirming the 
early indication from our C-COSMOS study of X-ray detected ETGs, we find 
significantly enhanced X-ray luminoaity in lower stellar mass ETGs 
(\(L_K\lesssim{10}^{11}L_{\astrosun}\)), relative to the local 
\(L_{X,gas}\propto L_K^{4.5}\) relation. The stacked spectra of these ETGs also 
suggest X-ray emission harder than expected from gaseous hot halos. This 
emission is consistent with inefficient accretion 
\({10}^{-5}-{10}^{-4}\dot{M}_{Edd}\) onto \(M_{BH}\sim 
{10}^{6}-{10}^{8}\,M_{\astrosun}\).
\end{abstract}


\section{Introduction}\label{sec:intro}

Diffuse X-ray emission from {normal} early type galaxies (ETG) 
{was first detected in the Virgo cluster with} the \textit{Einstein} 
X-ray Observatory {\citep{1979ApJ...234L..27F}. Although in a few cases 
(e.g., M86) the displacement of the X-ray emission from the stellar body of the 
galaxy pointed to hot gaseous halos, the} \(\sim\) arcminute resolution of 
{the} \textit{Einstein IPC} data {could not disentangle 
observationally} hot halos {from the populations of low-mass} X-ray 
{binaries} (LMXBs), {which was likely to  dominate the X-ray 
emission in a large number of the observed ETGs} \citep{1985ApJ...296..447T}. 
{Nonetheless}, several authors discussed how {the X-ray} emission 
and the inferred hot gas contribution could be used to constrain {both 
the binding mass of the galaxies and the physical evolution of the hot halos}, 
including feedback from SNIa and AGNs, and interaction with intracluster and 
group media \citep[see e.g.,][]{1989ARA&A..27...87F, 1987ApJ...312..503C, 
1991ApJ...376..380C, 1991ApJ...369..121D, 1991ApJ...367..476W, 
2003ARA&A..41..191M, 2007ApJ...665.1038C}.

Thanks to the high spatial resolution of \textit{Chandra} and imaging/spectral 
capabilities of the ACIS detector \citep{2003SPIE.4851...28G}, it has become 
possible to {directly detect and characterize} the different 
contributions to the X-ray emission of ETGs {in the near universe. With 
\textit{Chandra}, LMXB populations - both associated with Globular Clusters (GC) 
and in the stellar field of the ETG - have been studied} to distances of several 
tens of megaparsecs ({e.g., 
\citealt[][]{2000ApJ...544L.101S, 2002ApJ...574L...5K}; see review by} 
\citealt{2006ARA&A..44..323F}). {The integrated LMXB luminosity 
correlates well with the integrated ETG stellar luminosity} 
\citep[e.g.,][]{2004ApJ...611..846K, 2004MNRAS.349..146G}. {The scatter 
in this relation is a factor of 2-3, and it depends on both GC content 
\citep[]{2004ApJ...611..846K}, and the age of the stellar population 
\citep{2009ApJ...703..829K, 2010ApJ...721.1523K, 2011ApJ...729...12B, 
2014ApJ...789...52L}. These detailed high resolution studies} allowed 
\citet[][BKF11]{2011ApJ...729...12B} to accurately estimate the gaseous 
contribution to the X-ray luminosity \(L_{X,gas}\) of a sample of 30 
{ETGs within 32 Mpc. After subtracting LMXB, stellar, and nuclear 
contributions to the X-ray luminosity}, BKF11 found steep  scaling relations 
between \(L_{X,gas}\) and both the stellar \(L_K\) luminosity and the hot gas 
temperature \(T\). Subsequent work by Kim \& Fabbiano 
(\citeyear[][KF13]{2013ApJ...776..116K}; \citeyear[][KF15]{2015arXiv150400899K}) 
suggests that for \(L_{X,gas}>{10}^{40}\mbox{ erg}\mbox{ s}^{-1}\) the hot gas 
around ETGs is virialized, giving a new way to measure the total mass of these 
galaxies.

Beyond the local universe, \textit{Chandra} Deep Field studies (both detections 
and stacking of almost 1,000 ETGs) found that the mean \(L_X\) of \(L_B = 10^{10} 
L_{\astrosun}\) ETGs remains equal or mildly increases over \(z = 0-0.7\) 
\citep{2007ApJ...657..681L, 2008A&A...480..663T} and that the ratio of the soft X-ray 
luminosity to the B-band luminosity evolves mildly as \(L_{(0.5-2\mbox{ keV})}/ L_B 
\propto {(1 + z)}^{1.1 \pm 0.7}\) since \(z \approx 1.2\) 
\citep{2012MNRAS.422..494D}, suggesting  some heating mechanism preventing the 
hot gas from cooling, consistent with mechanical heating from radio AGNs. The 
evolution of the X-ray binary contribution to the X-ray emission (both of LMXBs 
and high mass binaries) across cosmic time, modeled by 
\citet{2013ApJ...764...41F}, led \citet{2013ApJ...776L..31F} to conclude that 
energy feedback from X-ray binaries is also important in the early universe.

{The COSMOS survey, covering \(\sim 2\) square degrees of sky with 
multi-wavelength coverage, which includes \textit{Chandra} X-ray coverage 
\citep{2007ApJS..172...38S, 2009ApJS..184..158E}, has increased the available 
sample of ETGs to over 6,600 galaxies \citet[][C14]{2014ApJ...790...16C}. This 
survey provides a good z coverage out to \(z\sim 2.5\) (and a few galaxies 
are detected out to \(z\sim 5\)).} C14 studied the sample of 69 X-ray detected 
ETGs in the \textit{Chandra}-COSMOS survey 
\citep[C-COSMOS,][]{2009ApJS..184..158E}, covering a redshift range \(0<z<1.5\), 
{to explore} redshift dependencies in the BKF11, KF13 scaling relations. 
In particular, C14 focused on the relation between \(L_{X,gas}\) and the K-band 
stellar luminosity (\(L_K\)) of ETGs {which is a good proxy of stellar mass for old stellar systems}. The latter has been evaluated in the local 
Universe by BKF11 as \(L_{X,gas}\propto {L_K}^{2.6 \pm 0.4}\) for sources with 
\(L_{X,gas} \leq 2 \times {10}^{41} \mbox{ erg}\mbox{ s}^{-1}\), while including 
sources with \(L_{X,gas}\) up to \(9 \times {10}^{42}\mbox{ erg}\mbox{ 
s}^{-1}\) KF13 found a steeper \(L_{X,gas}\propto {L_K}^{4.5 \pm 0.8}\) 
relation. C14 found that this latter relation also holds for X-ray detected ETGs 
with \(L_{X,gas} \leq \times {10}^{42} \mbox{ erg}\mbox{ s}^{-1}\), suggesting 
that the hot gas in these more distant galaxies may be virialized as well. 
However, more distant (\(z>0.9\)) and younger stellar age galaxies generally 
fall outside the relation observed in the local sample, and are X-ray 
overluminous with respect to their \(L_K\). For these sources C14 suggested that 
the higher X-ray luminosity may be due to the presence of hidden AGNs (as 
suggested by observed hardness ratios), merging phenomena, which are expected to 
increase the halo X-ray luminosity, or nuclear accretion \citep[that could also 
induce AGN activity during the merging phase,][]{2006ApJ...643..692C}.

In this work we perform a stacking analysis of the C-COSMOS ETG data in order to 
investigate the nature of fainter, X-ray undetected ETGs. We compare their 
properties with those of the sample of local ETGs investigated by BKF11 and 
KF13, and {with} the sample of ETGs detected in the C-COSMOS survey 
studied by C14. {This stacking study also allow us to further explore redshift dependencies. Our aim is to explore the properties of the non-stellar X-ray emission, i.e. the X-ray emission resulting from the subtraction of the expected X-ray binary contribution \citep[taking into account the cosmic evolution of the X-ray binary population][]{2013ApJ...776L..31F}. The resulting emission will probe both hot X-ray halos and the possible emission of active nuclei in these galaxies. In particular, the large number of low-stellar-mass ETGs in the sample will provide a unique probe of the existence of relatively low-mass massive black holes in their nuclei. These black holes should produce low-luminosity nuclear X-ray sources, if they are fueled by the ETG stellar outgassing \citep{2011ApJ...730..145V}.}

The paper is organized as follows: Section \ref{sec:sample} describes the sample 
selection and the data sets used in this work, Section \ref{sec:stack} is 
dedicated to the description of the adopted stacking procedure, in Section 
\ref{sec:discussion} we discuss our results, and Section \ref{sec:summary} is 
dedicated to our conclusions.In the following, we adopt the standard flat 
cosmology with \(\Omega_\Lambda = 0.73\) and \(H_0 = 70 \mbox{ km}\mbox{ 
s}^{-1}\mbox{ Mpc}^{-1}\).

\section{Sample Selection}\label{sec:sample}

The COSMOS survey \citep{2007ApJS..172...38S} provides a 
multi-wavelength characterization of over 1.5 million galaxies at redshift 
extending out to 5 in \(2\mbox{ deg}^2\) of the sky. The \textit{Chandra} X-ray 
coverage \citep[C-COSMOS][]{2009ApJS..184..158E} over \(0.9\mbox{ deg}^2\), 
gives us the means to study X-ray emission of ETGs, extending local 
studies to higher redshifts, and explore the redshift evolution of hot halos with large samples of galaxies.

Our ETG sample is selected from the most recent version of the COSMOS photometric catalog by \citet{2009ApJ...690.1236I}. The spectral energy distribution (SED) identification performed by \citet{2009ApJ...690.1236I} makes use of 7 elliptical galaxy templates with ages from 2 to 13 Gyr \citep{2007ApJ...663...81P}. The SED templates were generated with the stellar population synthesis package developed by \citet{2003MNRAS.344.1000B}, assuming an initial mass function from \citet{2003PASP..115..763C} and an exponentially declining star formation history. {We also collected additional} galaxy properties (stellar mass, age and star formation 
rate) from \citet{2010ApJ...709..644I}, and we made use of the most recent near-infrared photometry from the Ultra Deep Survey with the VISTA telescope \citep[Ultra-VISTA][]{2012A&A...544A.156M}\footnote{{Note that stellar mass reported in COSMOS photometric catalog, and obtained from the \citet{2003MNRAS.344.1000B} best-fit templates, do not incorporate the Ultra-VISTA K-band NIR data. Although this is only an issue if there is any residual star formation in the ETGs, \(L_K\) and \(M_{*}\) may not necessarily be completely equivalent.}}.

We selected galaxies classified as E or S0 and that have a reliable photometric 
redshift \citep{2009ApJ...690.1236I}. In addition we selected only sources 
with Ultra-VISTA detection both in J and K bands, since these are needed 
to evaluate the rest frame \(L_K\) (see below). This narrows our sample to 
10,808 (the COSMOS ETG sample considered in C14). We then selected only 
the 6436 ETGs falling in the C-COSMOS field, and then excluded the 48 
X-ray detected ETGs in common with the C-COSMOS identification catalog 
\citep{2012ApJS..201...30C}. This leads us to our final sample 
of 6388 C-COSMOS ETGs - not detected in X-rays - which constitutes the 
starting point of the stacking analysis described in this paper.
{Note that we selected sources with photometric redshift estimate in the catalog by 
\citet{2009ApJ...690.1236I}, while C14 selected their sources to have a 
spectroscopic or photometric redshift estimates in the C-COSMOS 
identification catalog, so the sample of ETGs selected by the latter 
authors (\(\sim 6600\) sources) is not fully contained in our sample.}

Following C14 we evaluated the K band luminosity of our sources from the 
Ultra-VISTA K-band aperture magnitude of the COSMOS photometric catalog. 
To calculate the rest frame K band luminosities we assumed a power law spectrum {(as appropriate over this small frequency range)}
\(f_\nu \propto \nu^{-\alpha}\) with \(\alpha = (J-K) /2.5 
\log{\left({\nu_J/\nu_K}\right)}\) (where J and K are the AB magnitudes from 
the COSMOS photometric catalog and \(\nu_J\) and \(\nu_K\) are the 
corresponding frequencies), and express the rest frame luminosities in solar 
luminosity as \(L_K = {10}^{-(K+\mbox{ext}-5.19)/2.5} \times {(1 + 
z)}^{\alpha-1} \times {(D_L/10)}^2 \), where \(z\) is the photometric 
redshift from the COSMOS catalog, \(D_L\) is the luminosity distance 
expressed in parsecs, and ext is the aperture correction parameter computed 
from the FWHM measured for all the COSMOS extended sources in the Hubble ACS 
data.

{We compared} the distributions of the physical parameters of the 6388 C-COSMOS ETGs 
{with} the entire 10,808 COSMOS ETG sample {finding no significant difference}, demonstrating {as expected} that the sample of ETGs not detected in X-ray is representative of COSMOS ETGs overall. Note that the luminosity threshold is a function of redshift because of the uniform magnitude limit of the Ultra-VISTA COSMOS survey (\(K_s \sim 24\), \citealt{2012A&A...544A.156M}). The specific star formation rate for the majority of the ETGs is in the regime where \citet{2010ApJ...709..644I} defines galaxies as quiescent, although higher star formation rate is observed in young ETGs. The properties (luminosity, mass, star formation rate and age) of the COSMOS ETGs are in agreement with those reported by \citep{2013A&A...558A..61M} for a sample of ETGs selected using multiple criteria (including color-color diagrams, spectra, star formation rate and SED classification). Masses and luminosities are in agreement with those of the K-selected elliptical galaxies of \citet{2014ApJ...783...25J}, extracted from a small area of the COSMOS field.

\section{Analysis and Results}\label{sec:stack}

Below we describe the procedures followed in our stacking analysis. The results are summarized in Tables \ref{tab:stack} and \ref{tab:stack_z}.

\subsection{Count Extraction at ETG Positions}

We used all 49 C-COSMOS observations to evaluate source and background 
counts for our sample of ETGs, excluding data in the ACIS-I chip gaps. We 
also excluded data from X-ray detected sources removing counts within a 10'' radius
of all sources in the C-COSMOS source identification catalog 
\citep{2012ApJS..201...30C}. We replaced the excluded counts by sampling the Poisson 
distribution of the pixel values nearby background regions with 
\textsc{dmfilth} \textsc{CIAO} task\footnote{\url{http://cxc.harvard.edu/ciao/ahelp/dmfilth.html}}. When a source lay in the field of view of more than 
one observation, we combined the data from the different observations 
using the prescriptions from \citet{2009ApJS..185..586P}, excluding 
observations where the source was more than 10 acrmin off-axis from the aim point.

\begin{figure}
\centering
\includegraphics[scale=0.5]{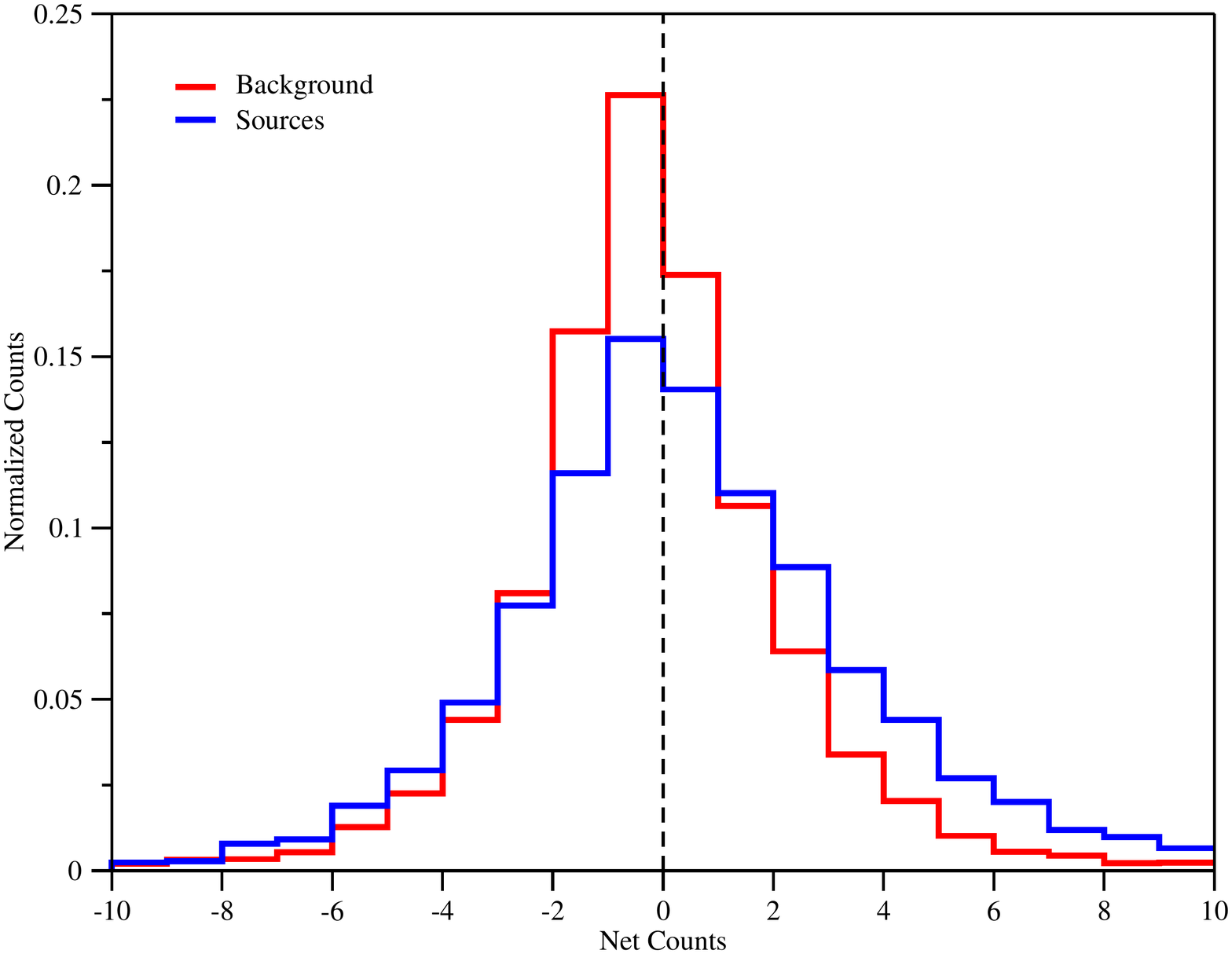}
\caption{Distribution of net counts for randomly distributed 
extraction regions (red line) and for the selected sample of sources (blue line).}\label{fig:background}
\end{figure}

Due to the variation of the \textit{Chandra} Point Spread Function (PSF) radius 
\(R_{PSF}\)\footnote{In the following we consider the \textit{Chandra}-ACIS radius enclosing 
90\% of the energy.}, we used an extraction radius \(r_{ext}\) that was determined by the 
source position in each observation in order to extract an optimal fraction of X-ray photons 
from the diffuse halo. For sources with morphological information available in the 
\textit{HST}-ACS catalog \citep{2007ApJS..172..219L} we adopted the source semi-major axes 
\(R\) as a measure of its physical extent. For sources with \(R>R_{PSF}\) we adopted values of 
\(r_{ext}\) that was the convolution of \(R\) and \(R_{PSF}\) {\(\sqrt{(R^2+R_{PSF}^2)}\)}; otherwise we adopted 
\(r_{ext}=R_{PSF}\) as we did for sources without morphological information in \textit{HST}-ACS catalog. {We notice however that sources with \(R>R_{PSF}\) represent only \(\sim 1\%\) of our sample, with extraction radii typically ranging between 1'' and 7''.} The average exposure time
for each source within the extraction circle, \(E\), was then evaluated. The histogram of 
extracted ETG position counts is shown in Figure \ref{fig:background} (blue line). {The small fraction of extraction regions with counts above \(\sim 5\) represent off-axis sources with large extraction radii.}

\subsection{Background Evaluation}

To evaluate the background we extracted counts from an annulus centered on 
the source coordinates, with inner-radius \(2r_{ext}\) and outer radius \(4r_{ext}\). Due 
to the very low signal, the background represents the largest source of 
uncertainty in this kind of analysis.
To estimate the systematic uncertainties introduced by the evaluation of the background, we compared the results of three different methods, obtaining consistent results. First we simply evaluated 
the background contribution to the total source counts from the mean count 
value of the pixels in the whole background annulus. Second, we adopted the clipping procedure of \citet[][method b]{2011ApJ...742L...8W}, which evaluates the mean count value \(\mu\) - and 
the standard deviation \(\sigma\) - of all pixels within a circle of radius \(4r_{ext}\) around each 
source position and then sets all pixels with count value 
larger than \(\mu+3\sigma\) to \(\mu\). This method suppresses the noise in the 
background annulus, but  also removes signal from the source\footnote{For the 
rare cases where \(\mu+3\sigma<1\), only pixels with two or more counts were 
clipped.}. Lastly, we divided the background annulus into eight 
\({45}\) degree sectors, and excluded all the sectors in which the mean count 
value was larger than \(\mu+3\sigma\), thus excluding regions 
of higher background without removing signal from the source.
{The} net counts \(N\) obtained with all three methods are consistent within statistics, {so in the following we adopted the clipping procedure of \citet[][method b]{2011ApJ...742L...8W}.}

To test the robustness of our results we also extracted counts from positions on the sky 
offset from the ETG coordinates by random distances between 15'' and 30'' both in right ascension and 
declination. The net 
counts of the non-source positions are shown in Figure \ref{fig:background}. They follow an 
approximatively Gaussian distribution centered on zero, as expected from background fluctuations, while the ETG counts show an excess of positive signal.
{To test if our results were dominated by a sub-population of ETGs we repeated the stacking procedure on sub-samples randomly drawn from the complete sample, progressively halving the subsample, obtaining results compatible with the full sample.}

\subsection{Stacking Procedure}\label{sec:stack}

To evaluate the count rate \(C\) for each stacked bin we stacked the observed source net counts {in 0.5-7 keV range} \(N\) and exposure \(E\) to obtain \(C=\sum_i{N_i}/\sum_i{E_i}\). From these counts we subtracted the expected contributions from stellar sources (LMXBs and Ultra Luminous X-ray sources, ULXs), and evaluated the average bin rest-frame X-ray luminosity as explained below. We corrected \(L_K\) to take into account the evolution of the stellar population (i.e., the fading effect, see C14). Note that, at a given redshift, younger galaxies have larger fading correction than older ones. To compare our results with those of KF13 and C14 we considered sources with \(9 < \log{(L_K)} < 13\).

\begin{figure}
\centering
\includegraphics[scale=0.8]{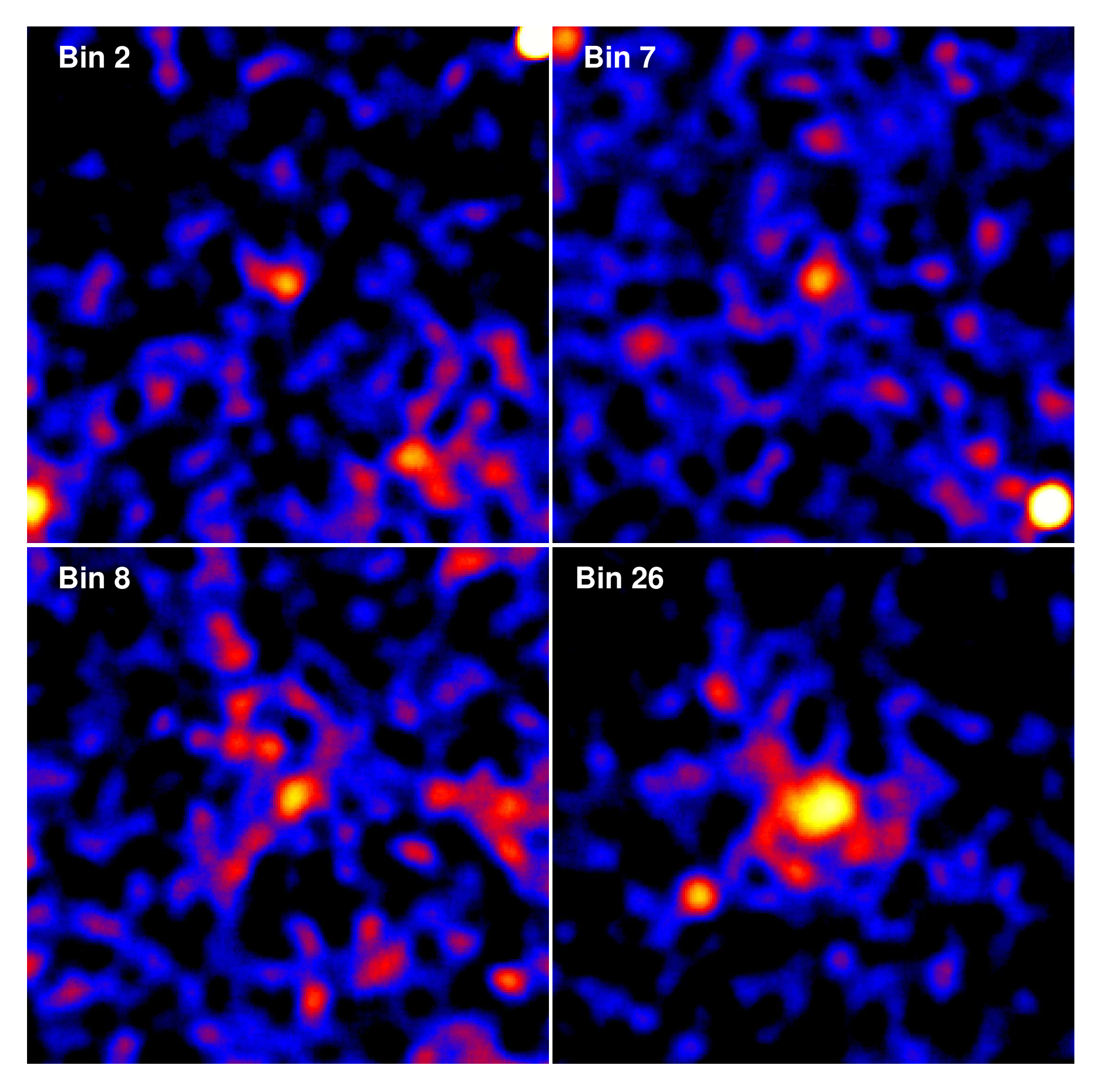}
\caption{Examples of 0.5-7 keV images of the \(L_K\) selected stacking bins.}\label{fig:bin_images}
\end{figure}

We constructed the stacks by binning in both \(L_K\) and \(z\), to ensure that the binning scheme did not produce any significant bias on the results. For the \(L_K\) binning, we first divided our sample in four \(\log{(L_K)}\) bins, \(9-10\), \(10-11\), \(11-12\) and \(12-13\). We then ordered the sources in each bin in increasing redshift, and progressively stacked them until we reached a minimum signal to noise ratio of 3; if this \(3\sigma\) threshold could not be reached we stacked the signal from all the remaining sources in the bin. Examples {of} 0.5-7 keV images {for} these stacking bins are presented in Figure \ref{fig:bin_images}. For the \(z\) binning, we followed the same procedure, dividing the sample into three bins \(0- 0.5\), \(0.5- 1\) and \(1- 1.5\), and then ordered the sources in each bin in increasing \(L_K\).

\subsection{Evaluation of Non-stellar \(L_X\) for Each Stacked Bin}\label{sec:lmxb}

To subtract the LMXB contribution from the total X-ray emission, we evaluated the average LMXB contribution to each bin using the 0.3-8 keV relation from 
\citet{2013ApJ...776L..31F}, as in C14:
\begin{equation}\label{eq:lmxb}
\log{\left({L_{LMXB}/M_{*}}\right)} = 30.259- 1.505 \times
\log{\left({\mbox{age}}\right)}- 0.421 \times 
\log^2{\left({\mbox{age}}\right)}+ 0.425 \times 
\log^3{\left({\mbox{age}}\right)}+ 0.135 \times 
\log^4{\left({\mbox{age}}\right)}\, ,
\end{equation}
where \(L_X\) is expressed in \(\mbox{erg}\mbox{ s}^{-1}\), \(M_{*}\) is the average stellar 
mass of the stacking bin expressed in solar masses and age is the average stellar age of the 
stacking bin expressed in Gyr. \footnote{{These average quantities, as well as their standard deviation in each bin is reported in Tables \ref{tab:stack} and \ref{tab:stack_z}.}} We also evaluated the contribution to X-ray luminosity from ULXs 
that could be present in the galaxies using the scaling relations from 
\citet{2007ApJ...666..870C} and \citet{2012MNRAS.419.2095M} between the ULX luminosity and the 
galaxy star formation rate. ULX contribution is more than an order of magnitude smaller 
than \(L_{X-LMXB}\) even for the least luminous \(L_K\) bins.

We converted the average \(0.3-8\) keV LMXB luminosities in each bin to count rates in the 
\(0.5-7\) keV interval using a power law model with index \(\Gamma=1.8\) (consistent with the 
typical spectrum of LMXBs) and Galactic absorption \(N_H=2.6 \times{10}^{20}\mbox{ cm}^{-2}\). 
For these parameters the conversion factor used from \textit{Chandra} count rates to fluxes is 
\(1.04\times {10}^{-11}\mbox{ erg}\mbox{ cm}^{-2}\mbox{ s}^{-1}\mbox{ 
cts}^{-1}\)\footnote{\url{http://cxc.harvard.edu/toolkit/pimms.jsp}}. We then subtracted these 
count rates from the stacked signals and converted back the resulting count rates into 
luminosities using a model comprising Galactic \(N_H\) plus a thermal \textsc{APEC} component 
with temperature depending on the source total X-ray luminosity. We adopted \(kT=0.7\), \(1\) and \(2\) keV for X-ray luminosities 
of \(L_X<{10}^{41}\), \({10}^{41}<L_X<{10}^{42}\) and \(L_X>{10}^{42} \mbox{ 
erg}\mbox{ s}^{-1}\), respectively, following BKF11 and 
\citet{2007ApJ...658..917D}. Finally, we K-corrected the luminosity to the 
\(0.3-8\) keV rest frame band. These LMXB-subtracted luminosities, that we indicate as \(L_{X-LMXB}\), are expected to include 
the emission from the hot gaseous halos, but may also include a possible contribution from any AGN harbored within these ETGs.

\begin{figure}
\centering
\includegraphics[scale=0.45]{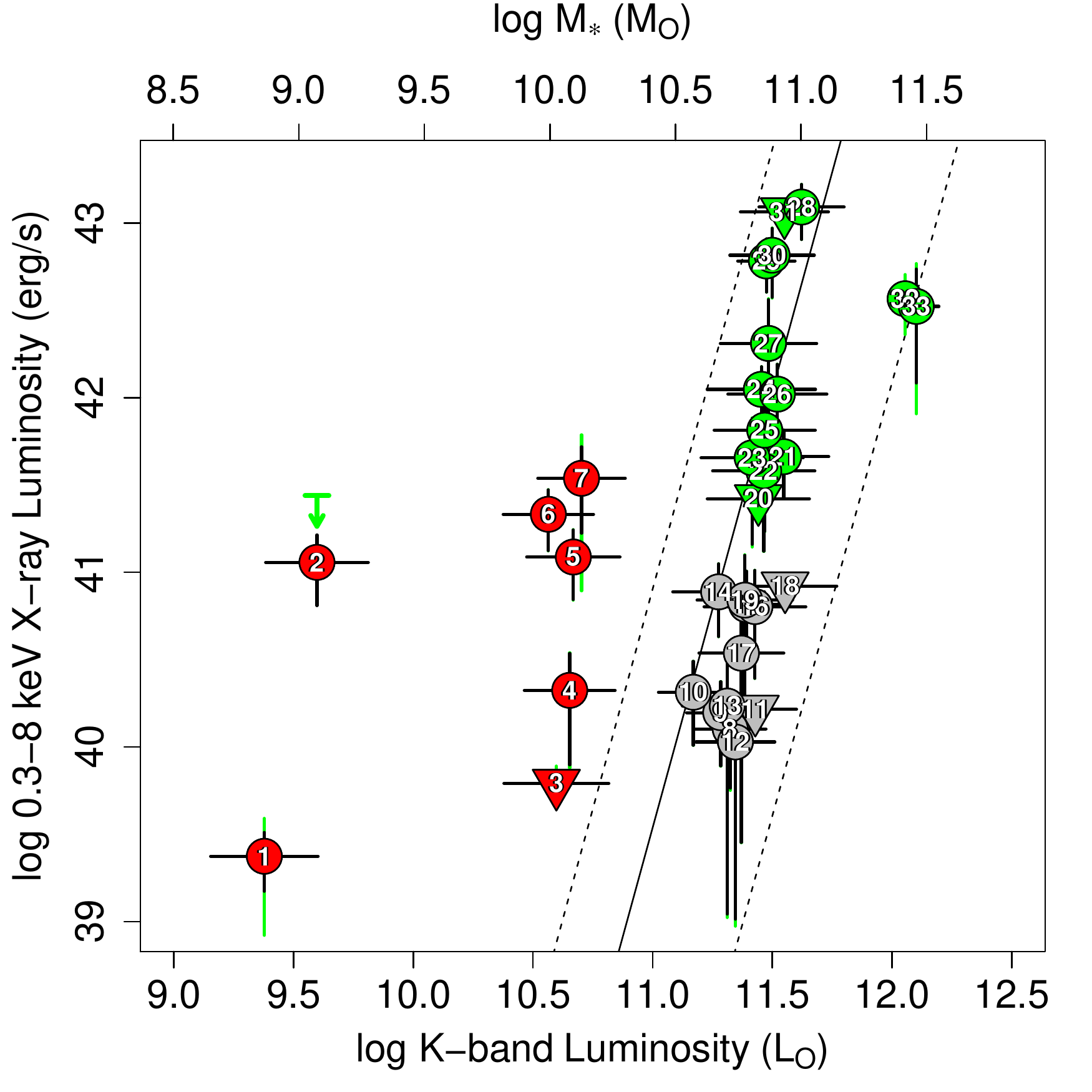}
\includegraphics[scale=0.45]{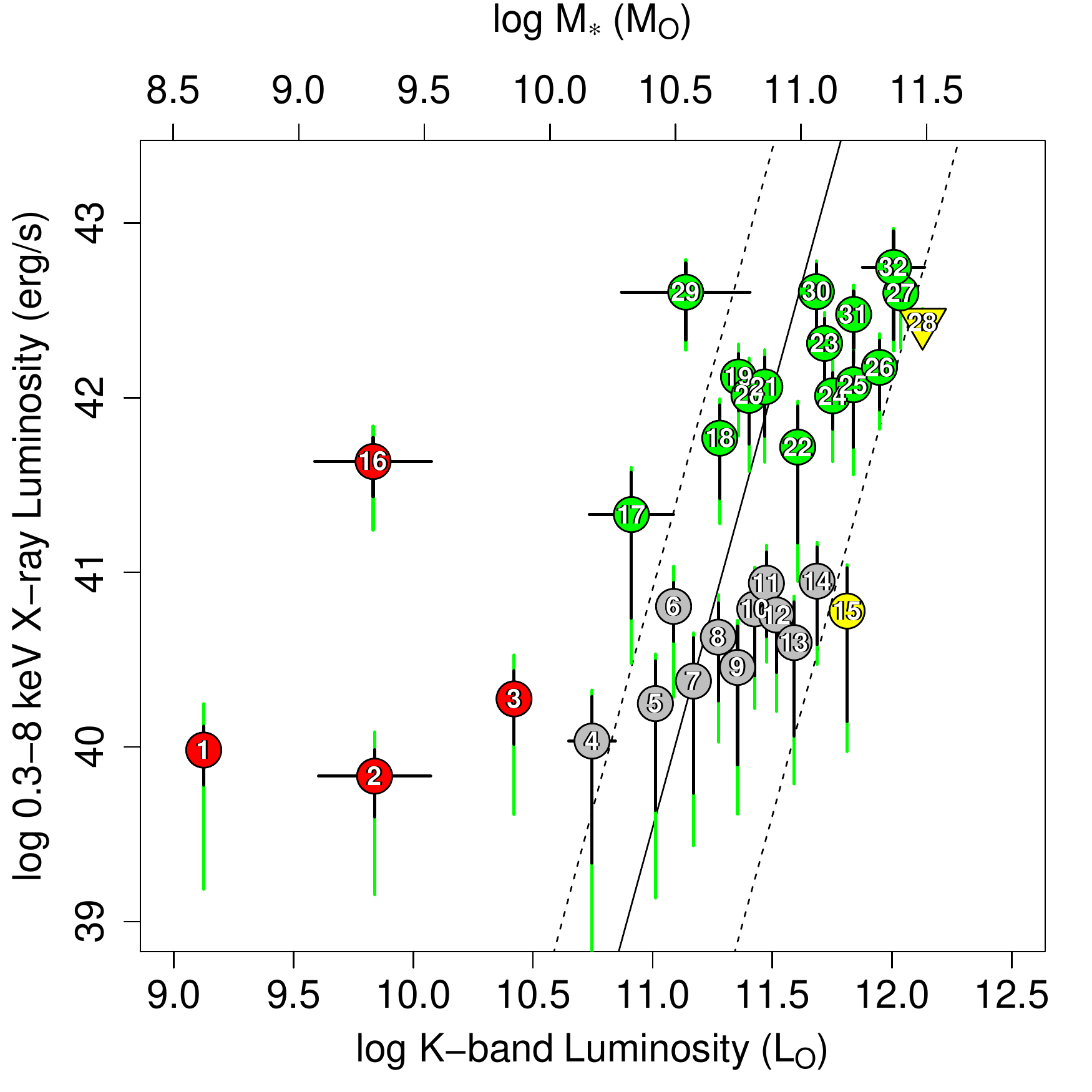}
\caption{X-ray luminosity (after LMXBs subtraction) versus K-band luminosity for the stacking bins selected in \(L_K\) ({Left Panel}) and redshift ({Right Panel}). The horizontal and vertical {black} error bars indicate the statistical uncertainties on \(L_K\) and \(L_{X-LMXB}\), respectively, {while the vertical green error bars show the uncertainty on \(L_{X-LMXB}\) due to redshift spread in each bin.} {The stellar mass \(M_*\) from COSMOS catalog is plotted on the top axis.} The local relation by KF13 is represented with a full black line. The dashed lines are drawn to include the most \(L_{X,gas}\) luminous source ETG (M87) and the most \(L_{K}\) luminous ETG (NGC1316) in the Local Sample (BKF11, KF13), respectively. Bins are labelled as in Table \ref{tab:stack} (for the left panel) and Table \ref{tab:stack_z} (for the right panel): red, X-ray over-luminous with respect to the local \(L_{X,gas}-L_K\) relation; yellow, X-ray under-luminous bins; grey and green bins following the local relation with \(L_{X-LMXB}\) lower and higher than \({10}^{41}\mbox{ erg}\mbox{ s}^{-1}\), respectively. Triangles pointing down represent upper limits.}\label{fig:stack}
\end{figure}

Figure \ref{fig:stack} shows the results in the \(L_{X-LMXB}-L_K\) plane for both \(L_K\) and \(z\) binning (Left and Right panels, respectively), {where} on the upper axes we plot \(M_*\) as taken from COSMOS catalog. In both binning schemes the overall scatter plots are consistent, showing that we do not suffer from strong binning biases. By construction, the \(L_K\)-first binning provides information on the extreme redshift bins (both high and low), while the \(z\)-first binning covers more fully the \(L_K\) range. The bins are numbered in Figure \ref{fig:stack}, following Tables \ref{tab:stack} and \ref{tab:stack_z}. We use different colors to visually divide the stacking bins into four groups: red, for bins X-ray over-luminous with respect to the local \(L_{X-LMXB}\) and \(L_K\) relation (BFK11, KF13, see C14); yellow, for X-ray under-luminous bins; and grey and green for bins following the local relation with \(L_{X-LMXB}\) lower and higher than \({10}^{41}\mbox{ erg} \mbox{ s}^{-1}\), respectively.

{As noted in Section \ref{sec:intro}, Eq. \ref{eq:lmxb} has an observed 
spread of a factor 2-3, which can be related to both the presence of GC LMXB in 
the ETGs and to the spread in age of the stellar population 
(\citealt{2004ApJ...611..846K, 2010ApJ...721.1523K}; BKF11; 
\citealt{2014ApJ...789...52L}). Given the parameter space of Figure 
\ref{fig:stack}, this spread will not change our conclusions.}

\subsection{Hardness Ratios}

Hardness ratios (HRs) were evaluated to characterize the ETG average spectral properties. 
\(HR=(H-S)/(H+S)\), where \(S\) is the soft \(0.5-2\mbox{ keV}\) counts and \(H\) the hard 
\(2-7\mbox{ keV}\) counts. These HRs are those of the LMXB-subtracted emission.
The contributions of these sources were evaluated and subtracted by applying the procedure in 
{Section} \ref{sec:lmxb} to each energy band. {We evaluated the errors on HRs as 
due to both uncertainties on counts and to uncertainties on estimate of LMXB contribution due 
to spread of stellar age and mass in the bins.}

\begin{figure}
\centering
\includegraphics[scale=0.45]{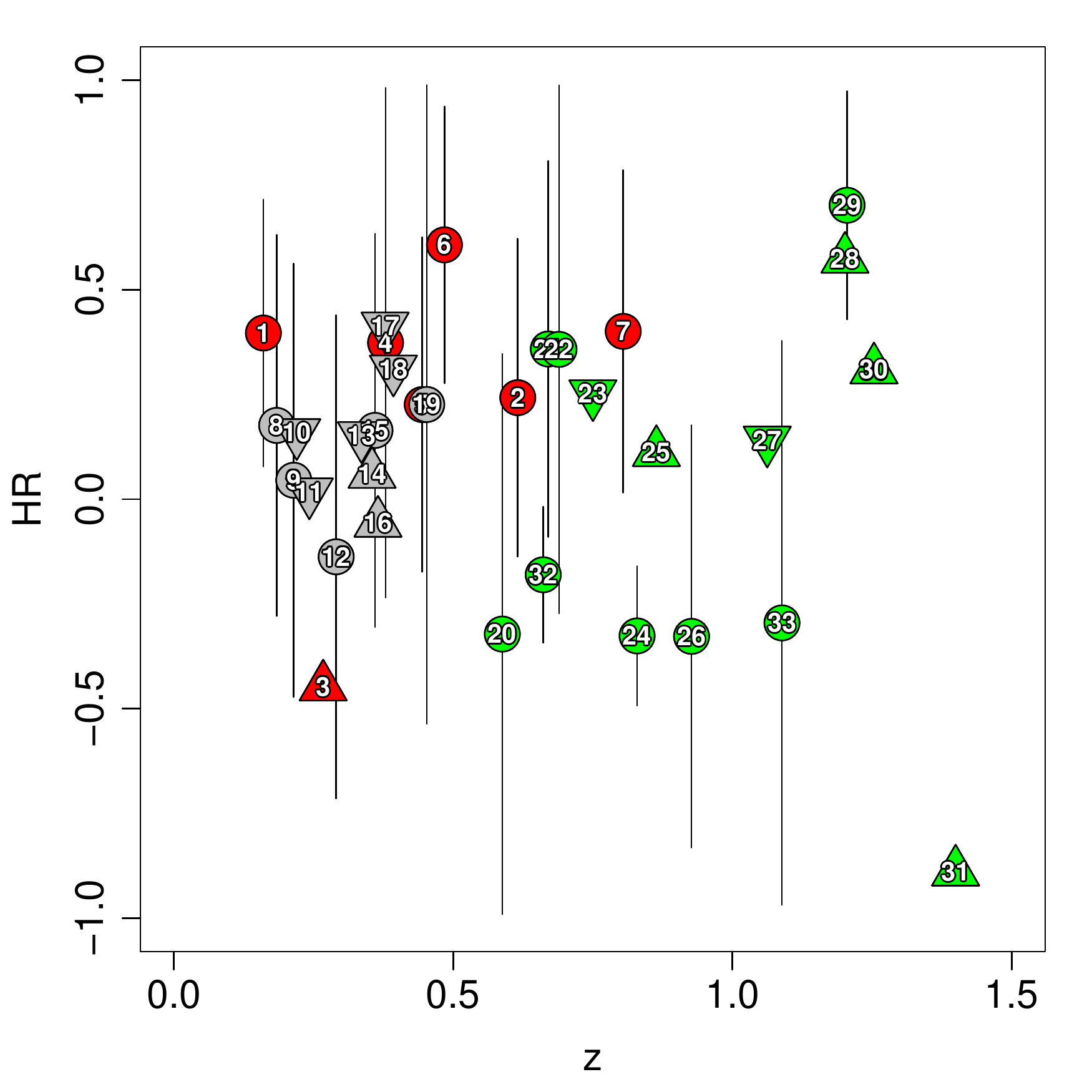}  
\includegraphics[scale=0.45]{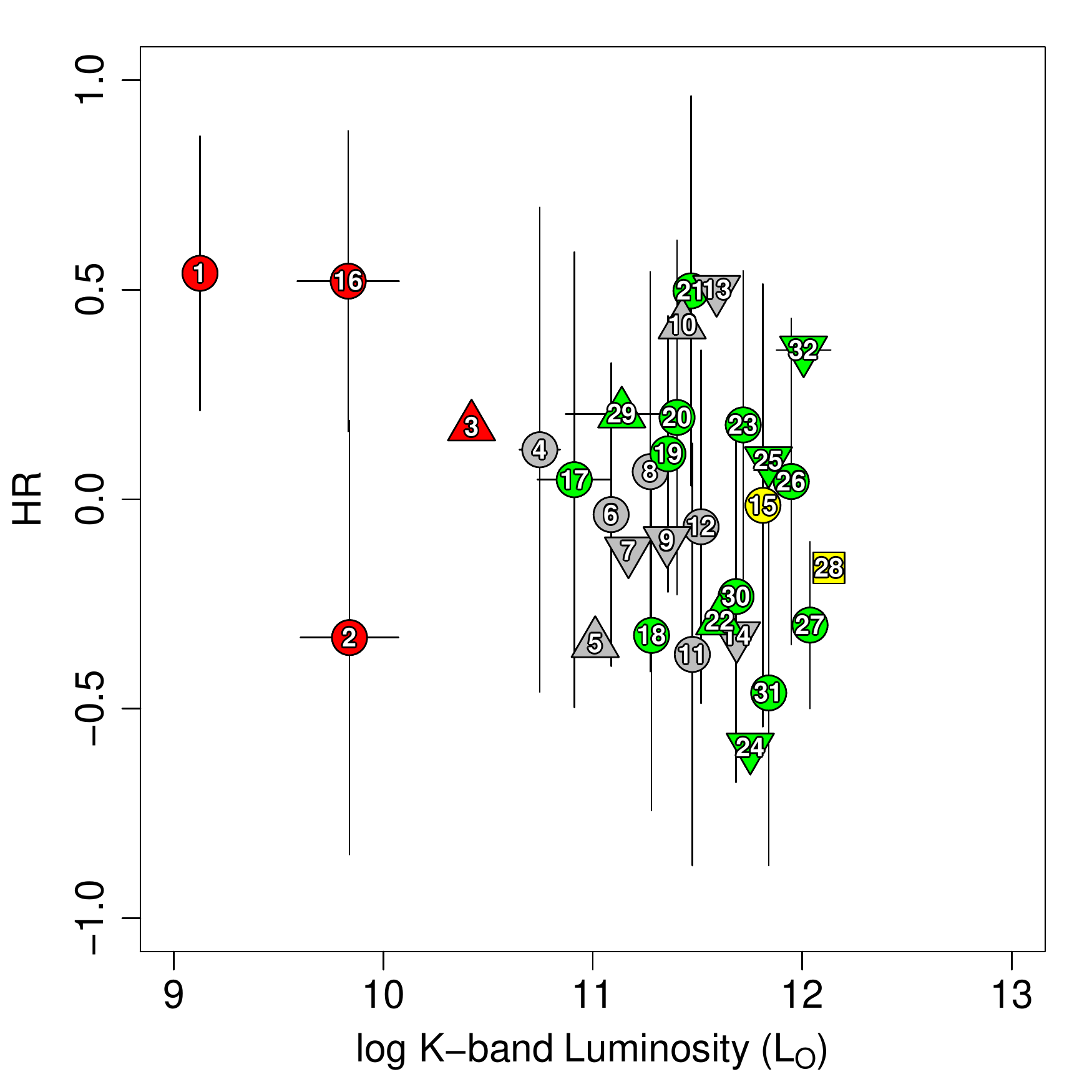}  
\caption{Left: HRs of the \(L_K\)-first stacking bins versus redshift. Right: HRs of the 
z-first stacking bins versus \(L_K\). Triangles pointing up and down represent lower and upper 
limits on HR, respectively{, while the box for bin 28 represents unconstrained HR.}}\label{fig:hr}
\end{figure}

In Figure \ref{fig:hr} we show the HRs for the stacking bins using the same color code as in Figure \ref{fig:stack}. We plot HR versus \(z\) in the Left panel, using the \(L_K\)-first binning, which gives a better \(z\) sampling. We plot HR versus \(L_K\) in the Right panel, using the \(z\)-first binning, to explore the \(L_K\) behavior. The left panel of Figure \ref{fig:hr} shows a distinct clustering in HR-\(z\) space of bins from separate loci of the \(L_{X-LMXB}-L_K\) plane, as shown by the clusters of different color points. The Right panel shows that ETGs with larger stellar mass (\(L_K\)) tend to have lower HR. Given the scaling relations of ETGs (BKF11; KF13; KF15), this could just reflect a relation with total binding mass, with more massive galaxies being more efficient at retaining large hot gaseous halos.

\section{Discussion}\label{sec:discussion}

\begin{figure}
\centering
\includegraphics[scale=0.45]{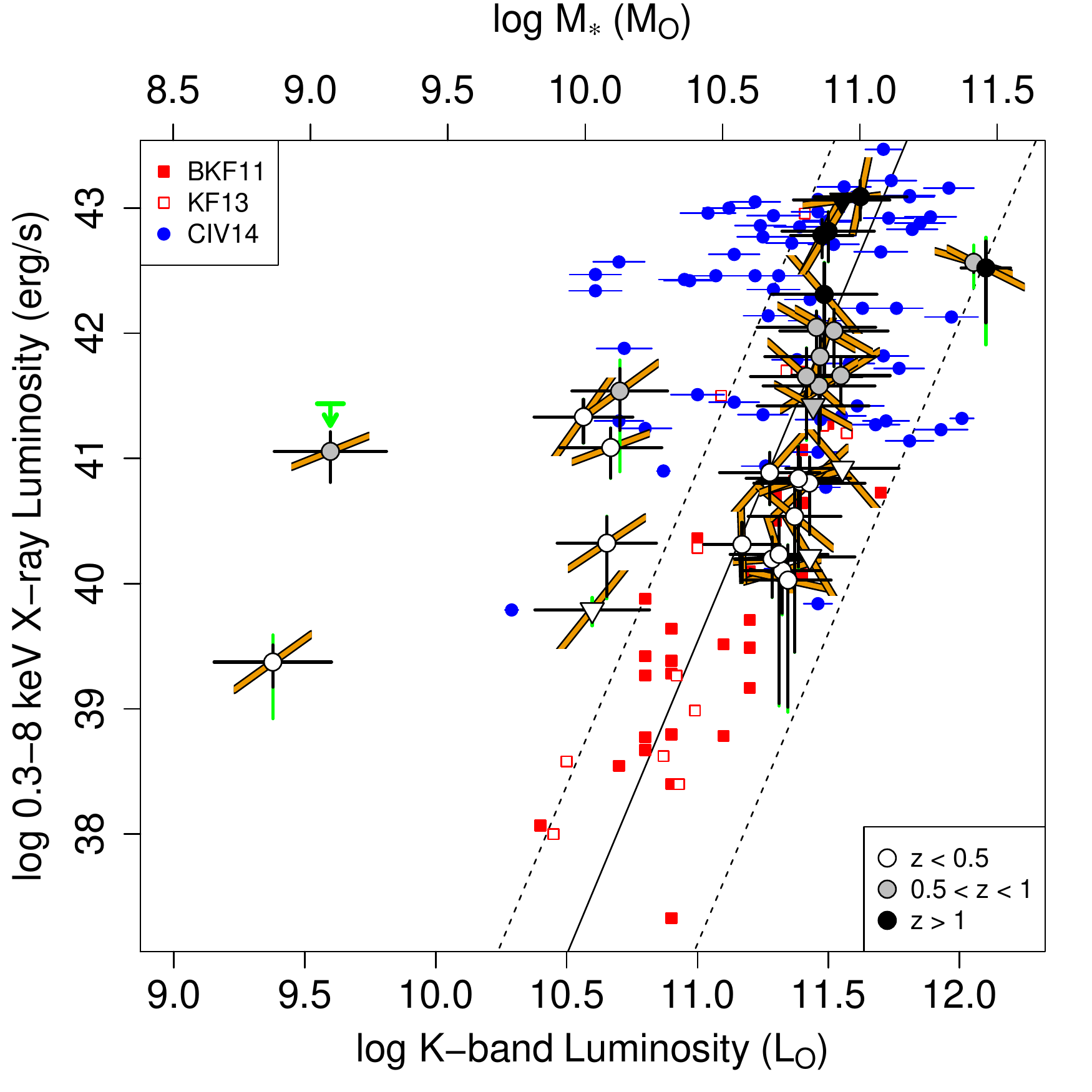}
\includegraphics[scale=0.45]{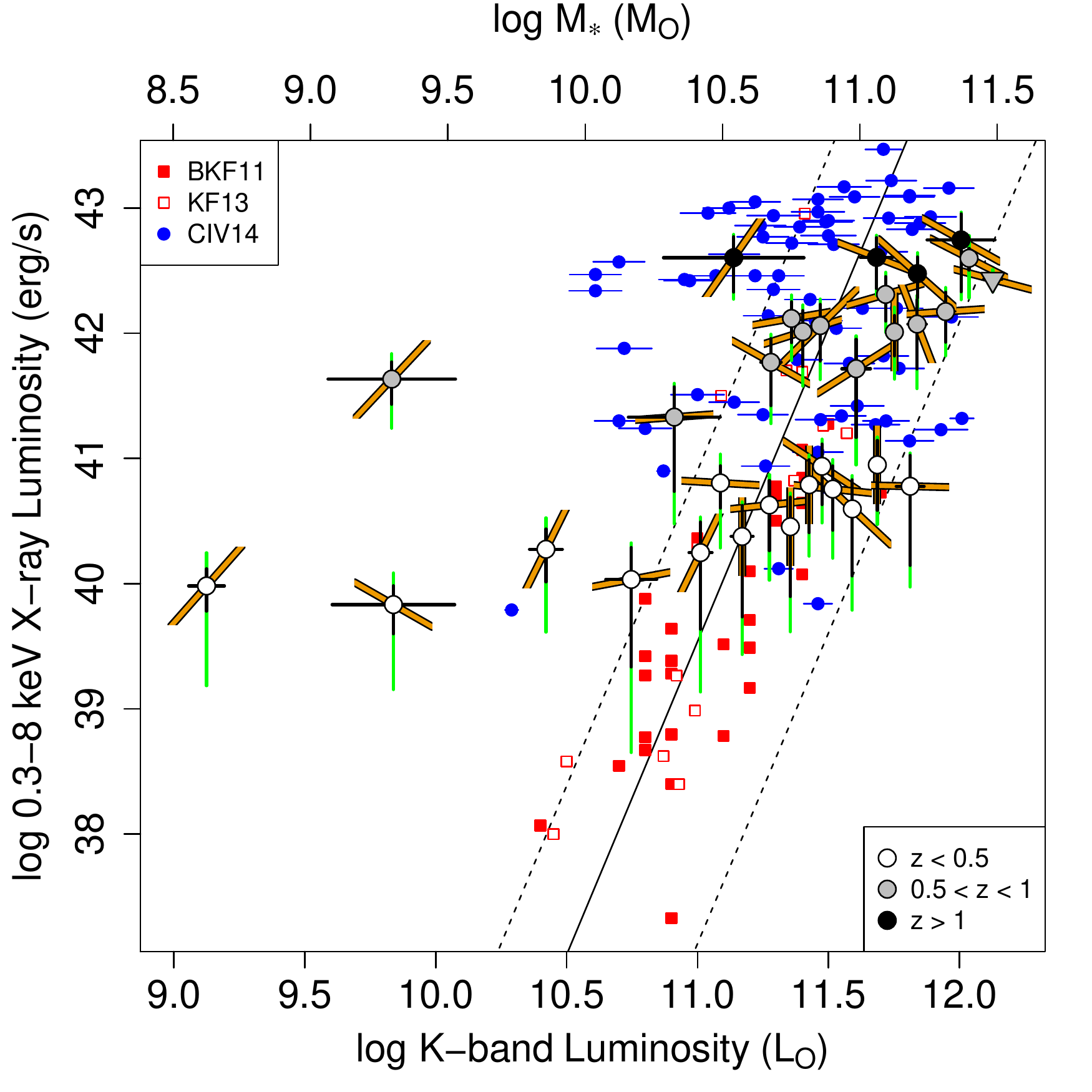}
\caption{Same as Figure \ref{fig:stack}, but with bins labelled according to their average redshift, and HRs indicated by the slope of the orange segments. Black error bars reflect statistical uncertainties, vertical green bars show the uncertainty on \(L_{X-LMXB}\) due to redshift spread in each bin. Red boxes show the BKF11 and KF13 samples, blue circles show the C14 sample of C-COSMOS X-ray detected ETGs. Triangles pointing down represent upper limits.}\label{fig:stack_hr}
\end{figure}

Figure \ref{fig:stack_hr} summarizes our results in the \(L_{X-LMXB}-L_K\) plane and compares them with the COSMOS ETG X-ray detections (C14) and with the near universe ETG data (BKF11, KF13). We show the local KF13 relation with a solid line, and the scatter in this relation with the two dashed lines. These lines are chosen to pass through the points for the most \(L_{X,gas}\) luminous source (M87) and the most \(L_{K}\) luminous source (NGC1316), respectively. The COSMOS stacked-ETG results are overall consistent with the C14 COSMOS detections, but significantly increase the coverage at lower {total} stellar luminosities{ / masses}. While most of the detected signals are compatible with the local KF13 relation, there are several detections that lie both to the left and to the right side of it, that is, bins that appear to be respectively over and under-luminous in X-ray with respect to their \(L_K\) {or \(M_*\)}.

In Figure \ref{fig:stack_hr} we also show the hardness ratios for each stacked bin with lines with slopes proportional to \(HR\). We find positive HRs (i.e., harder spectra) in stacked bins outside the dashed line on the left. These are the lower \(L_K\) galaxies with significant X-ray excess (the red points in Figure \ref{fig:hr}, see also C14). We also find hard \(HR > 0\) in some points with \(L_{X-LMXB}> {10}^{41}\mbox{ erg}\mbox{ s}^{-1}\) and \(z>1\) (the three top green points in Figure \ref{fig:hr}), to the left of the solid line.

Instead, lower HR ratios (i.e., softer spectra) are found in bins following the local 
relation (between the two dashed lines). This is especially so for those to the right of the 
solid line. This suggests that they have a larger than average hot gas content up to 
\(z\sim 1\).

\begin{figure}
\centering
\includegraphics[scale=0.45]{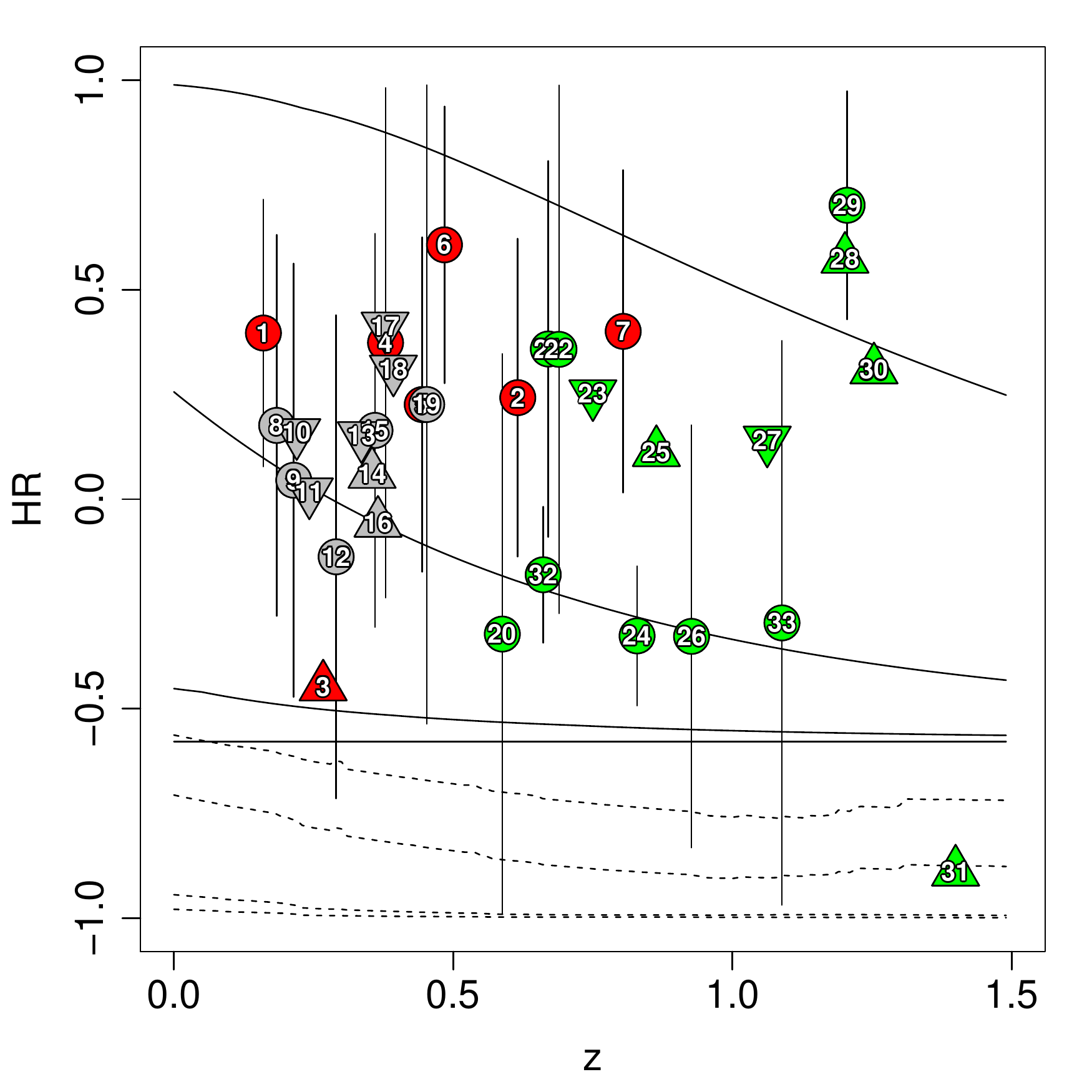}    
\includegraphics[scale=0.45]{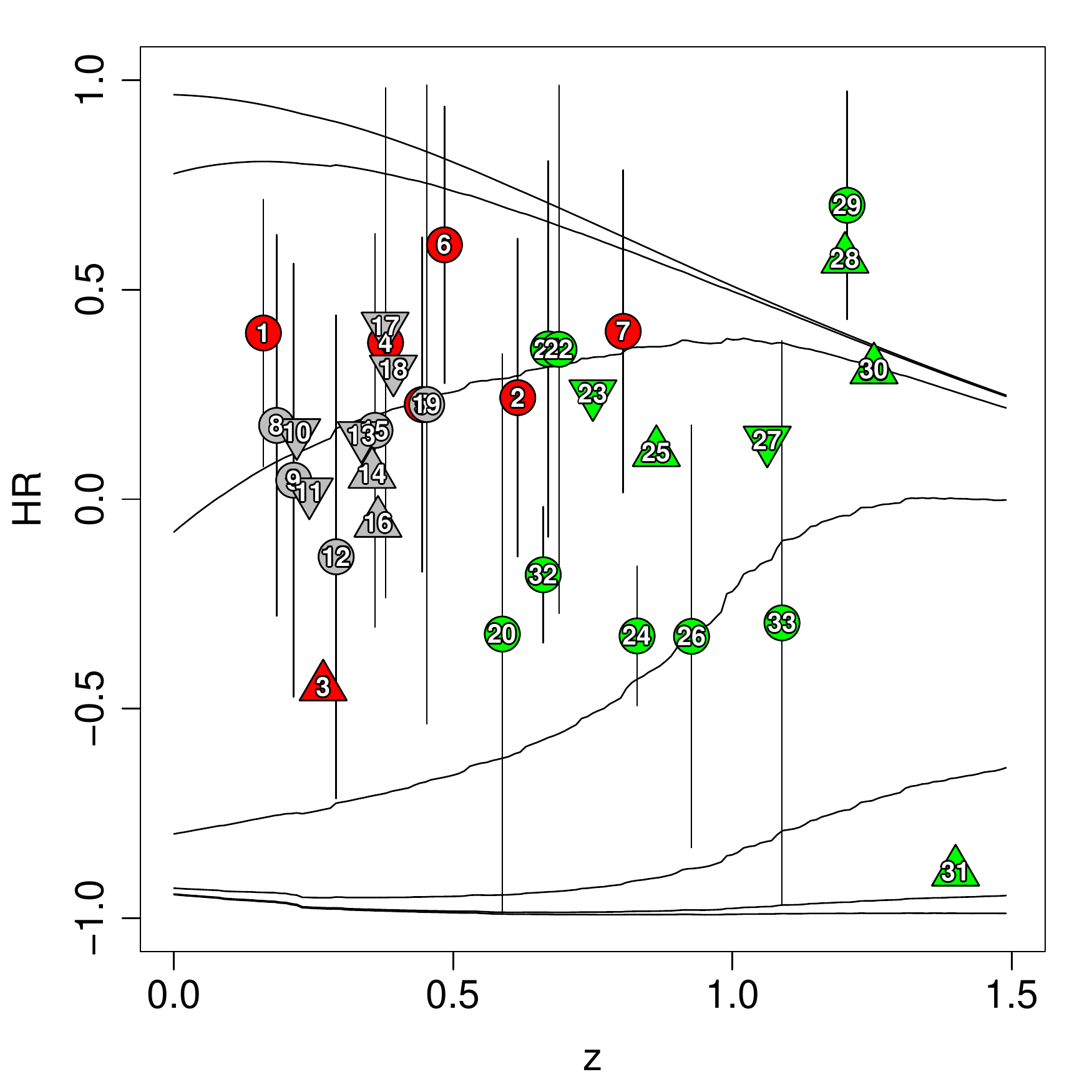}
\caption{Left: HR-z scatter diagram of \(L_K\)-first bin. {The vertical error bars indicate uncertainties on \(L_{X-LMXB}\) due to counts and to LMXB subtraction.} Triangles pointing up and down represent lower and upper limits on HR, respectively. {Squares represent unconstrained HRs}. Dashed lines show HRs expected from thermal models (\textsc{APEC}) with increasing {temperatures from} bottom to top 0.7, 1, 2 and 3 keV. The full lines represent simulated HR from power-law models with fixed slope \(\Gamma=2\) and increasing intrinsic absorption form bottom to top 0, \({10}^{21}\), \({10}^{22}\) and \({10}^{23}\) \(\mbox{cm}^{-2}\). Right: Same as left panel, but for models comprising a power-law {PL} with fixed slope \(\Gamma=2\) and intrinsic absorption of \({10}^{23}\) \(\mbox{cm}^{-2}\), plus a thermal model with temperature of 1 keV {(APEC)}. The relative normalization of the two models is, from {bottom to top}, \({10}^{-3}\), \({10}^{-2}\), \({10}^{-1}\), \(1\), \({10}\), \({10}^{2}\) and \({10}^{3}\), corresponding to a contribution of the power-law component to the total flux of \(\sim 0.2\%\), \(\sim 2\%\), \(\sim 15\%\), \(\sim 62\%\), \(\sim 94\%\), \(\sim 99\%\) and \( 100\%\), respectively.}\label{fig:hr_pl_therm}
\end{figure}

Figure \ref{fig:hr_pl_therm} compares the HR-z scatter diagram with a set of 
spectral models. On the Left, we show the expected HRs for thermal \textsc{APEC} 
(dashed lines; from bottom to top, \(kT= 0.7\), \(1\), \(2\) and \(3\) keV) and 
power-law models (solid lines; fixed slope \(\Gamma= 2\) and increasing intrinsic absorption 
from bottom to top, \(N_H=0\), \({10}^{21}\), \({10}^{22}\) and \({10}^{23}\) 
\(\mbox{cm}^{-2}\)). The \textsc{APEC} models are the expected emission of thermal gaseous 
emission, while the power-law models are typical AGN spectra. To illustrate cases of mixed 
emission from a hot gas component and a heavily obscured AGN, on the right panel of Figure 
\ref{fig:hr_pl_therm} we plot composite models consisting of \(kT=1\) keV \textsc{APEC} 
emission with line of 
sight \(N_H\) plus power-law models with slope \(\Gamma=2\) and intrinsic absorption 
\(N_H={10}^{23}\) \(\mbox{cm}^{-2}\). From the bottom to the top the relative normalization of 
the two models varies, to reflect an increasing contribution of the power-law model to the 
total flux, ranging from \(\sim 0.2\%\) to \(\sim 100\%\).

The X-ray over-luminous bins (red points) have generally high HRs 
consistent with being dominated by AGN emission, either with ``normal" AGNs with absorbing columns \(\gtrsim {10}^{22}\mbox{ cm}^{-2}\), or in the case of the composite gas + AGN emission even highly obscured Compton Thick AGNs (\(N_H > {10}^{24}\mbox{ cm}^{-2}\), \citealt{1999ApJ...522..157R,2006ApJ...648..111L,2010ApJ...724..559L}).

The bins following the local relation with \(L_{X-LMXB}<{10}^{41}\mbox{ erg}\mbox{ s}^{-1}\) (grey points) show {\(HR \sim  (-0.5) - 0.3\), lower than those of} the over-luminous X-ray bins. 
{A soft HR first suggests emission from hot gas. However, Figure \ref{fig:hr_pl_therm} shows that pure APEC thermal models with solar metallicity always have \(HR < -0.6\). Hence these objects have HR values inconsistent with pure gaseous emission.}

These HRs {can instead be} explained {with combinations} of thermal and AGN emission. Figure \ref{fig:hr_pl_therm} Right {shows models with varied ratios of APEC and power-law emission. In this model the observed HRs imply AGN components accounting  for \(15\% - 60\%\)} of the X-ray luminosity. Given the X-ray luminosities of this {sub-sample} (Figures \ref{fig:stack}, \ref{fig:stack_hr}), these ratios {then} imply AGN luminosities ranging from \(\sim 5\times {10}^{39}\mbox{ erg} \mbox{ s}^{-1}\) to \(\sim 5\times {10}^{40}\mbox{ erg} \mbox{ s}^{-1}\). This range is in agreement with the low-luminosity \textit{Chandra} X-ray sources in the near universe sample (BKF11, {\citealt{2005ApJ...624..155P}}), which have luminosities ranging from \(\sim {10}^{38}\mbox{ erg} \mbox{ s}^{-1}\) to \(\sim 5\times {10}^{41}\mbox{ erg} \mbox{ s}^{-1}\).

The bins with \(L_{X-LMXB} > {10}^{41}\mbox{ erg}\mbox{ s}^{-1}\) (green points) have hard \(HR>0\) incompatible with a low absorption power-law model (e.g., bins 21, 22 and 25). The three highest redshift bins in this group (e.g., bins 28, 29 and 30) have even harder \(HR>0.5\) that are compatible only with absorbing columns larger than \({10}^{23}\mbox{ cm}^{-2}\). This suggests that these galaxies harbor (on average) highly obscured Compton Thick AGNs that dominate their X-ray luminosity.

In summary, our stacking analysis suggests that on average AGN emission is likely to be present in ETGs. These AGNs may be faint and hidden in hot gas dominated ETGs, which are consistent with the near Universe scaling relation of virialized halos (KF13, KF15, see C14). {In contrast}, AGN emission appears to dominate the non-stellar X-ray emission in two classes of ETGs: (1) low-stellar mass ETGs (\(\log{(L_K/L_{\astrosun})} \lesssim 10.5\)), and (2) higher redshift ETGs (\(z>0.5\), \(>1\)) especially those overluminous in X-rays relative to the near universe \(L_{X-LMXB}-L_K\) scaling relation. The latter group {has} HR values suggesting highly absorbed Compton Thick AGNs. We note that these AGNs were not visible in the COSMOS colors and spectra, since AGNs were excluded {in} our ETG sample {selection process} (C14).

\subsection{Accretion Regimes for the AGNs}\label{sec:accretion}

\begin{figure}
\centering
\includegraphics[scale=0.45]{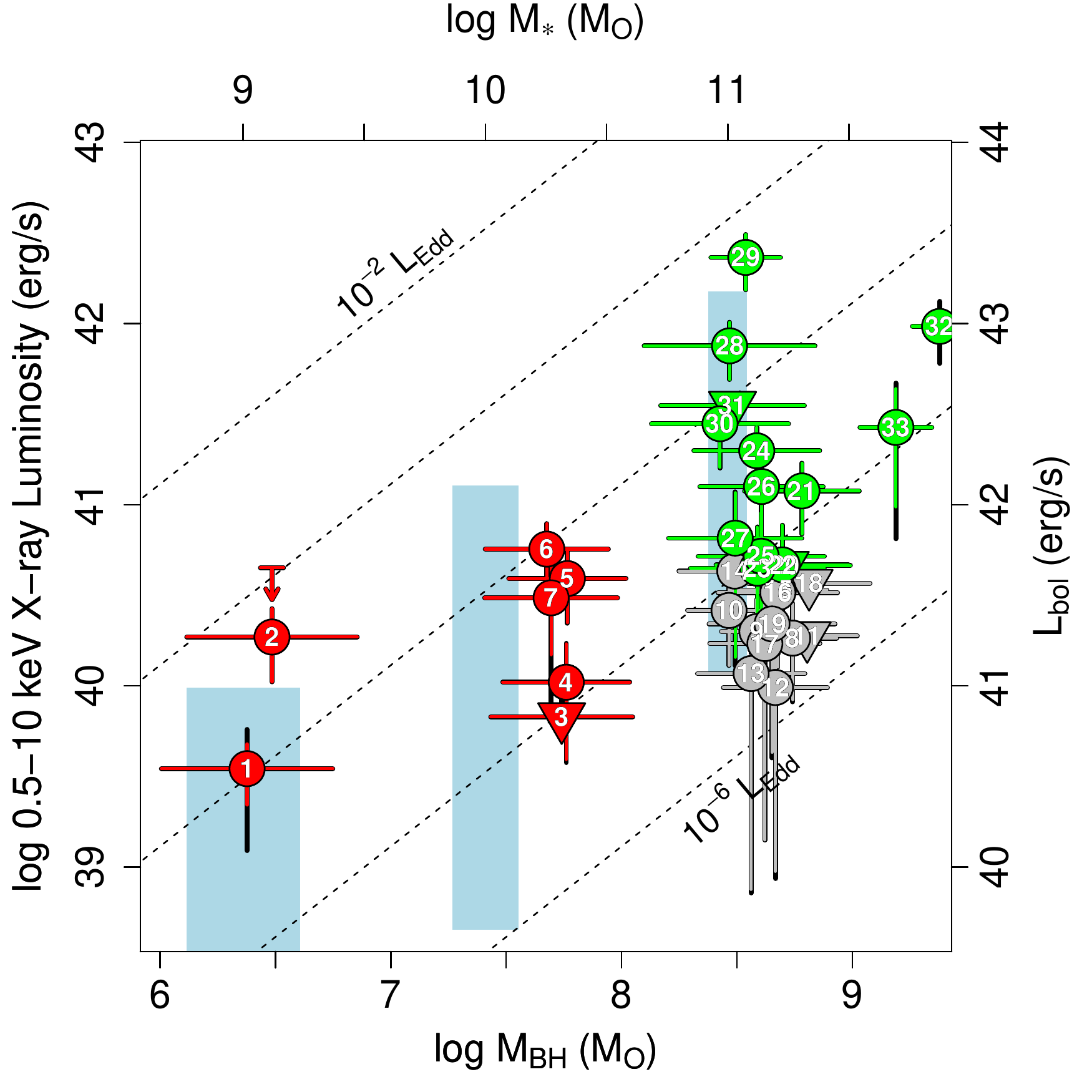}
\caption{X-ray luminosity (after LMXBs subtraction) versus black hole mass 
\(M_{BH}\) (evaluated from \(M_*\) according to the {\citet{2013ARA&A..51..511K}} relation) for the \(L_K\)-first stacking bins. The stellar mass \(M_*\) from COSMOS catalog is plotted on the top axis. Lines of constant \(L_{bol}/L_{Edd}\), given \(M_{BH}\) and \(L_{bol}\), are shown as diagonals (from bottom to top \({10}^{-6}\), \({10}^{-5}\), \({10}^{-4}\), \({10}^{-3}\) and \({10}^{-2}\)). The bins are color-labeled as in Figure \ref{fig:stack}. Light blue boxes represent the predictions of the \citet{2011ApJ...730..145V} models for \(M_{*}={10}^9\, M_{\astrosun}\), \({10}^{10}\, M_{\astrosun}\) and \({10}^{11}\, M_{\astrosun}\), where the top of each box refers to radiatively efficient accretion flows and the bottom refers to RIAFs. 
}\label{fig:volonteri}
\end{figure}

Figure \ref{fig:volonteri} shows the results of the \(L_K\)-first stacking, with \(L_{X-LMXB}\) 
plotted against stellar mass \(M_*\) and the mass of the nuclear black hole, \(M_{BH}\), {obtained} from the \(M_{BH} - M_*\) 
relation of {\citet{2013ARA&A..51..511K}. Evaluating \(M_{BH}\) from the \(M_{BH} - M_*\) relation \citet{2013ApJ...764..184M} would yield black hole masses between \(0.1\) and \(0.3\) dex smaller than the former. Similarly}, one could derive \(M_{BH}\) from the \(M_{BH} - L_K\) relation from \citet{2007MNRAS.379..711G}, {obtaining black hole masses} \(\sim 0.5\) dex larger than the former. Uncertainties in these estimate are typically of a factor of {\(\sim 2-3\)}
\citep{2002ApJ...574..740T,2007MNRAS.379..711G} and evolution to \(z\sim 1-2\) may provide a 
systematic black hole mass increase of a similar factor \citep{2010ApJ...708..137M}. These 
effects will not change our conclusions in the {3-dex} wide range covered by Figure 
\ref{fig:volonteri}.

On the right hand {axes} we plot \(L_{bol}=10\,L_{X-LMXB}\), an appropriate bolometric correction for {low-luminosity} AGNs that radiate at a few \(10\)s \(\%\) of \(L_{Edd}\) \citep{2006ApJ...648..128K,2010MNRAS.402.2637S}. Given this nominal \(L_{bol}\) we can plot 
diagonal lines of constant \(L_{bol}/L_{Edd}\). These lines range from \({10}^{-2}\) to 
\({10}^{-6}\) {\(L_{bol}/L_{Edd}\)}. 

We see that the massive high \(L_{X-LMXB}\) galaxies (green points) and the low-mass X-ray overluminous (red points) galaxies have implied Eddington ratios in the \({10}^{-3}\) to \({10}^{-5}\) range. RIAFs are generally believed to apply to all black hole accretion below \({10}^{-2} L/L_{Edd}\), so we would expect all these sources to be RIAFs. If the bolometric correction for RIAFs is smaller than {that considered here}, as may well be the case \citep{1995ApJ...452..710N}, {the \(L_{bol}\) values on the y-axis should be correspondingly lower (as the Eddington ratios on the currently plotted dashed lines).}

Alternatively the true \(L_{bol}\) of the three high \(z\) bins of massive high \(L_{X-LMXB}\) galaxies (green points 28, 29, 30) could be greatly underestimated. The HRs of these bins also imply highly absorbed emission (\(N_H > {10}^{23}\mbox{ cm}^{-2}\), Figure \ref{fig:hr_pl_therm}). If these are then Compton Thick AGN they may have much larger intrinsic luminosities, with factors of 100 being quite plausible \citep[e.g.,][]{1999ApJ...522..157R,2006ApJ...648..111L}. They would have intrinsic luminosities \(L_{0.5-10\mbox{ keV}} \sim {10}^{45}\mbox{ erg}\mbox{ s}^{-1}\) and \(L_{bol}/L_{Edd}\) in the \(\sim{10}^{-2}\) range, so that they could lie within the normal AGN regime. 

{To investigate this possibility, we built SEDs of ETGs in our sample using optical-infrared magnitudes reported in COSMOS photometric catalog, since a ``hidden" AGN may give rise to an IR excess with respect to an average ETG SED. We then fitted these SEDs with the same elliptical galaxy template used by \citet{2009ApJ...690.1236I} without finding any significant deviation. To confirm this, we evaluated the \(L_{2-10-LMXB}\) luminosity of the stacking bins assuming a power-law spectrum (as appropriate for AGN spectra) and compared that with their IR luminosity. In all cases the X-ray luminosities are at least 3 order of magnitudes lower than the IR luminosities, even assuming absorbing columns \(\sim{10}^{23}\mbox{ cm}^{-2}\), indicating that at IR wavelength a putative AGN component is expected to be significantly smaller than the galactic emission.}

The galaxies {that are} consistent with the local \(L_K-L_{X-LMXB}\) gaseous relation (grey points), with X-ray colors suggesting thermal emission plus a possible AGN component, have smaller Eddington ratios \(\sim{10}^{-5}\) to \({10}^{-6}\) (these ratios of course assume that the entire emission is due to an AGN; as shown in Figure \ref{fig:hr_pl_therm}, the AGN component may range from \(\sim 15-60\%\)). 

We also plot in Figure \ref{fig:volonteri} the predictions from \citet{2011ApJ...730..145V} of 
\(L_{X, AGN}\) for three values of \(M_*\) for both efficient and radiatively inefficient 
accretion flows \citep[RIAFs][]{1995ApJ...452..710N}. These authors' models use the stellar density profile scaling with stellar mass, and predict 
the accretion rate from the normal stellar mass loss due to stellar evolution, and the 
consequent massive black hole luminosity for various radiative efficiencies\footnote{We note 
that in \citet{2011ApJ...730..145V} models assume for ETGs a stellar age of 12 Gyr. We 
corrected for the average age \({10}^{9.5}\) Gyr of C-COSMOS ETGs, 
which yields slightly smaller emitted X-ray luminosities (see Eq. 9 in 
\citealt{2011ApJ...730..145V} ).}. The X-ray luminosity coming from accretion is computed as 
\(L_{X, AGN}/L_{Edd} = \eta_X \eta \dot m\),  where \(\eta_X = 0.1\) is the X-ray fraction of 
the bolometric luminosity, and \(\dot{m}=0.1\dot{M}/\dot M_{Edd}\) is the accretion rate in 
Eddington units with a \(10\%\) accretion efficiency. Radiatively efficient accretion flows 
have \(\eta =1\), while inefficient ones have \(\eta=\dot {m}/\dot {m}_{cr}\), and for them 
$\dot m<\dot m_{cr}=3\times 10^{-2}$. 

In order to compare our results with \citet{2011ApJ...730..145V} models we converted the 
{\(0.5-7\mbox{ keV}\) LMXB subtracted counts into \(0.5-10\mbox{ keV}\) band luminosities, \(L_{0.5-10\mbox{ keV}}\) assuming a pow-law spectrum as appropriate for AGN spectra}. Based on these models, inefficient 
accretion into the nuclei could explain even all the \(L_{X-LMXB}\) of the massive, high 
\(L_{X-LMXB}\) ETGs (green points), except for the most luminous ones, which are likely to be 
Compton thick AGNs (see above). The low-mass X-ray overluminous ETGs (red points) are also 
consistent with the \citeauthor{2011ApJ...730..145V} model. Considering that a large fraction 
of the X-ray emission is due to hot gas (see above), the grey points show instead a possible 
AGN emission even lower than predicted for radiatively inefficient accretion. In these ETGs, 
the \(L_{X-LMXB}\) is consistent with the local \(L_{X,gas}-L_K\) relation. 

This result can have the following possible explanations, that are not mutually exclusive.
\begin{enumerate}
\item The Volonteri et al.'s models assume that the whole of their estimated mass 
accretion rate (coming from the instantaneous mass loss of all stars, and obtained by 
integrating over the whole stellar density profile)  indeed reaches  the massive black hole, 
while this represents an upper limit on the mass accretion rate of the massive black hole (some 
gas may be outflowing, or may  reach the massive black hole with a delay, or may possess 
angular momentum and form a disk, etc.).
\item The accreting mass, that has already entered the accretion radius, is reduced on its way 
to the massive black hole, due to a wind from a RIAF 
\citep{1995ApJ...452..710N,1999MNRAS.303L...1B}; similarly, close to the nucleus, there may be 
a heating mechanism for the accreting gas, as a nuclear jet, again impeding or lowering the 
accretion rate \citep[e.g.][]{2012ApJ...758...94P,2011ApJ...737...23Y}.
\item Accretion onto the black hole is (temporarily) not taking place, as may be the case 
during feedback-regulated activity cycles. Numerical simulations show that, during these 
cycles, AGN outbursts are followed by major degassing of the circumnuclear region, with a 
precipitous drop of the nuclear accretion rate; the outbursts are separated by long time 
intervals during which the galaxy is replenished again by gas from the stellar mass losses, 
until a new nuclear outburst takes place \citep{2010ApJ...717..708C}.
\item \(\eta_X\) might be lower than 1 as assumed here (see also 
\citealt{2011ApJ...730..145V}), or that \(\eta\) at low \(\dot{m}\) decreases  faster than 
adopted in the Volonteri et al.'s models, so that lower \(L_{X, AGN}\) should be predicted 
\citep[e.g.,][]{2014ARA&A..52..529Y}.
\end{enumerate}

Our results extend to higher X-ray luminosities a recent analysis of ETGs in the AMUSE survey of Virgo and field galaxies with \textit{Chandra} \citep{2015ApJ...799...98M}. AMUSE results indicate a significant presence of supermassive black holes in low stellar mass galaxies. Due to the lower redshifts of the AMUSE ETGs, that survey is sensitive to X-ray luminosities (\(\sim 1.3\times{10}^{38}\mbox{ erg}/\mbox{s}\)) one order of magnitude fainter are than our faintest bin 1 (\(\sim 2.4\times{10}^{39}\mbox{ erg}/\mbox{s}\)). For these dim sources the stellar wind accretion proposed by \citet{2011ApJ...730..145V} would predict higher X-ray luminosities than those observed, \citeauthor{2015ApJ...799...98M} conclude that these black holes are probably powered by inefficient advection dominated accretion flows or, alternatively, by an outflow/jet component. We found that low-luminosity, X-ray ``overluminous" ETGs can be adequately described by the RIAF fueled by stellar winds as proposed by \citet{2011ApJ...730..145V}. Instead, ETGs with \(L_{X-LMXB}>{10}^{42}\mbox{ erg}\mbox{ s}^{-1}\) may host Compton Thick AGNs.
{Similarly, when comparing with the results of \citet{2007ApJ...657..681L} and \citet{2012MNRAS.422..494D} on \textit{Chandra} Deep Fields, we extend to higher X-ray luminosities and to lower optical luminosities (that is, lower stellar masses), finding sources that significantly exceed the expected emission from hot-gas dominated ETGs, with \(\log{(L_{0.5-2\mbox{ keV}} / L_B) / (\mbox{erg}/\mbox{s} / L_{\astrosun})} > 30 \) (see Figure 7 in \citealt{2012MNRAS.422..494D}).}

We also note that, as reported by \citet{2015A&A...574L..10C}, the revision of black hole scaling relations by \citet{2013ARA&A..51..511K} indicate that the local mass density in black holes should be five times higher than previous estimates. These authors advance the possibility that a sizeable population of highly obscured AGNs with covering
fraction of obscuring material \(\sim 4\pi\) would explain the revised
upward value of the SMBH local mass density without exceeding the limits imposed by the X-ray and IR background. An example of these obscured AGNs can be that observed in the ETG ESO565-G019 with \textit{Suzaku} and \textit{Swift}/BAT \citep{2013ApJ...773...51G}.

{Finally, we note that the hard X-ray slopes of the more luminous ETG bins could be due to contamination from X-ray jets. However, the combined 1.4GHz VLA COSMOS Large and Deep Project catalogs \citep{2010ApJS..188..384S} offer radio counterparts only to \(\sim 2\%\) of our sample of ETGs. The presence of radio emitters among non X-ray detected ETGs will be addressed after the release of the JVLA-COSMOS Project data.}

\section{Summary and Conclusions}\label{sec:summary}

We have followed up our previous work (C14) investigating the X-ray / K-band properties of the 69 ETGs detected in the Chandra COSMOS survey \citep{2009ApJS..184..158E}, with a luminosity and hardness ratio stacking analysis of the \textit{Chandra} data for the sample of 6388 early type galaxies (ETGs) not detected in X-rays. The stacking 
analysis allows the investigation of the average features of individually X-ray undetected 
ETGs. Our ETG sample was selected as representative of normal galaxies, with no evidence of AGN emission in their spectra and multi-band photometry (C14, Section 2). Our purpose was to investigate the properties of the X-ray emission from hot gas and from possible ``hidden" AGN components that may be revealed by their X-ray emission. To this purpose, the expected contribution of {LMXBs} was subtracted from the X-ray luminosity of each stacked bin, using established scaling relations with the K-band luminosity (BKF11). All luminosities were corrected to the local rest frame; X-ray hardness ratios were based on the observed counts (after subtraction of the LMXB contribution), and were compared to emission models for the appropriate average bin redshift.

This analysis has led to the following conclusions:

\begin{enumerate}
\item {On average AGN emission may be present in all ETGs.}
\item {AGNs may be faint and hidden in hot gas dominated ETGs, which are consistent with the local universe scaling relation of virialized halos (KF13, KF15, see C14).}
\item {AGN emission appears to dominate the non-stellar X-ray emission in
low-stellar mass ETGs (\(\log{(L_K)} \lesssim 10.5\) in solar units), which show marked X-ray luminosity excesses relative to their K-band luminosity (a proxy of stellar mass), when compared with the local ETG \(L_{X,Gas} – L_K\) relation of the local universe (BKF11). This result confirms the conclusions of C14 and extends them to lower  stellar mass galaxies.}
\item {AGN emission is also prominent in higher redshift ETGs (\(z>0.5\), \(>1\)) especially those with X-ray luminosity \(\gtrsim {10}^{42}\mbox{ erg}\mbox{ s}^{-1}\). The latter group have HR values suggesting highly absorbed Compton Thick AGNs. We note that these AGNs were not visible in the COSMOS colors and spectra, since AGNs were excluded by our ETG sample (C14).}
\item {Given the nuclear BH mass predicted by the \(M_{BH}-L_K\) relation \citep{2013ARA&A..51..511K}, the hidden AGNs in the X-ray over-luminous low stellar mass ETGs are consistent with the presence of radiatively inefficient accretion fueled by stellar outgassing \citep{2011ApJ...730..145V}. The highly absorbed higher \(L_{X-LMXB}\) AGNs hidden in massive, higher z ETGs instead suggest efficient accretion in highly obscured nuclei.}
\end{enumerate}

\acknowledgments
This work was partially supported by NASA contract NAS8-03060 (CXC), and 
NASA Chandra grant G01-12110X. SP acknowledges financial support from MIUR 
grant PRIN 2010-2011, project `The Chemical and Dynamical Evolution of the 
Milky Way and Local Group Galaxies', prot. 2010LY5N2T. FC acknowledges 
financial support by the NASA contract 11-ADAP11-0218. AP thanks Andy 
Goulding for useful discussion and suggestions. 
{ME and GF thank  the Aspen Center for Physics and the NSF Grant \#1066293 for hospitality during the completion of this paper.}
This research has made use of 
software provided by the Chandra X-ray Center (CXC) in the application 
packages CIAO, ChIPS, and Sherpa. This research is based on observations 
collected at the European Organisation for Astronomical Research in the 
Southern Hemisphere, Chile, program 179.A-2005 (UltraVISTA survey).

\newpage

\begin{table}
\begin{center}
\caption{Main features of the \(L_K\) selected stacking bins: (1) bin number; (2) number of stacked sources; (3) average rest frame \(L_K\) in the bin (with {corresponding} standard deviation); 
(4) average redshift in the bin (with {corresponding} standard deviation); 
{(5) average stellar age in the bin (with {corresponding} standard deviation); (6) average stellar mass in the bin (with {corresponding} standard deviation)}; (7) signal to noise ration of the stacked signal; (8) average 0.3-8 X-ray luminosity after LMXB subtraction {estimated from a thermal model. The first uncertainty reported in parenthesis is due to standard deviation of counts, stellar age and stellar mass in the bin, while the second one takes into account the additional standard deviation of redshift.}; (9) hardness ratio (with error {due to standard deviation of counts, stellar age and stellar mass in the bin})}\label{tab:stack}
\resizebox{\textwidth}{!}{
\begin{tabular}{ccccccccc}
\hline
\hline
(1) bin number & (2) number of sources & (3) \(<{\log({L_K/L_{\astrosun}})}>\) & (4) \(<{z}>\) & (5) \(<{\log({\mbox{age}/\mbox{yr}})}>\) & (6) \(<{\log({M_*/M_{\astrosun}})}>\) & (7) \(S/N\) & (8) \(<L_{X-LMXB,\,0.3-8}/({10}^{40}\mbox{ erg}\mbox{ s}^{-1})>\) & (9) HR \\
\hline
1 & 93    & 9.38(0.22)  & 0.16(0.04) & 9.66(0.26)  & 8.93(0.27)  & 3.12 & 0.24(0.09,0.13)       & 0.40(0.32)  \\
2 & 946   & 9.60(0.21)  & 0.62(0.34) & 9.43(0.35)  & 8.98(0.38)  & 2.60 & 11.40(5.07,10.72)     & 0.24(0.38)  \\
3 & 272   & 10.60(0.22) & 0.27(0.07) & 9.60(0.24)  & 10.07(0.33) & 3.09 & \(<0.62\)             & \(>-0.45\)  \\
4 & 190   & 10.65(0.19) & 0.38(0.02) & 9.50(0.21)  & 10.11(0.30) & 3.00 & 2.11(1.45,0.26)       & 0.37(0.61)  \\
5 & 41    & 10.67(0.19) & 0.44(0.01) & 9.43(0.19)  & 10.11(0.32) & 3.01 & 12.23(5.64,0.73)      & 0.23(0.40)  \\
6 & 43    & 10.56(0.19) & 0.48(0.01) & 9.38(0.21)  & 10.06(0.25) & 3.24 & 21.48(8.62,0.67)      & 0.61(0.33)  \\
7 & 1017  & 10.70(0.18) & 0.80(0.19) & 9.31(0.18)  & 9.99(0.47)  & 3.03 & 34.53(19.32,8.27)     & 0.40(0.39)  \\
8 & 4     & 11.32(0.15) & 0.18(0.01) & 10.01(0.03) & 11.01(0.17) & 3.14 & 1.27(0.77,0.15)       & 0.18(0.45)  \\
9 & 7     & 11.28(0.14) & 0.21(0.01) & 9.96(0.10)  & 10.89(0.13) & 3.12 & 1.57(0.87,0.06)       & 0.05(0.52)  \\
10 & 7    & 11.17(0.15) & 0.22(0.01) & 9.86(0.23)  & 10.78(0.15) & 3.02 & 2.06(1.12,0.01)       & \(<0.16\)   \\
11 & 20   & 11.43(0.17) & 0.24(0.02) & 9.94(0.09)  & 11.05(0.20) & 3.24 & \(<1.64\)             & \(<0.02\)	  \\
12 & 43   & 11.35(0.16) & 0.29(0.02) & 9.88(0.10)  & 10.93(0.21) & 3.06 & 1.07(0.96,0.13)       & -0.14(0.58) \\
13 & 59   & 11.31(0.19) & 0.34(0.01) & 9.75(0.19)  & 10.84(0.22) & 3.21 & 1.71(1.60,0.13)       & \(<0.15\)   \\
14 & 13   & 11.28(0.19) & 0.35(0.01) & 9.73(0.16)  & 10.78(0.21) & 3.12 & 7.73(3.44,0.18)       & \(>0.06\)	  \\
15 & 13   & 11.39(0.16) & 0.36(0.01) & 9.75(0.16)  & 10.94(0.22) & 3.04 & 6.53(3.61,0.01)       & 0.16(0.47)  \\
16 & 16   & 11.43(0.21) & 0.37(0.01) & 9.72(0.22)  & 10.91(0.25) & 3.04 & 6.36(3.89,0.17)       & \(>-0.06\)  \\
17 & 45   & 11.37(0.18) & 0.38(0.01) & 9.68(0.22)  & 10.88(0.23) & 3.05 & 3.44(3.16,0.08)       & \(<0.41\)	  \\
18 & 45   & 11.55(0.22) & 0.39(0.01) & 9.70(0.19)  & 11.02(0.28) & 3.03 & \(<8.34\)             & \(<0.31\)	  \\
19 & 91   & 11.39(0.20) & 0.45(0.03) & 9.69(0.22)  & 10.89(0.24) & 3.09 & 6.92(5.57,1.03)       & 0.23(0.76)  \\
20 & 216  & 11.44(0.21) & 0.59(0.06) & 9.61(0.25)  & 10.92(0.28) & 3.04 & \(<26.36\)            & -0.32(0.67) \\
21 & 27   & 11.55(0.19) & 0.67(0.01) & 9.57(0.23)  & 11.00(0.28) & 3.30 & 45.92(19.33,0.08)     & 0.36(0.45)  \\
22 & 161  & 11.46(0.21) & 0.69(0.01) & 9.56(0.23)  & 10.91(0.29) & 3.06 & 38.05(24.83,1.38)     & 0.36(0.63)  \\
23 & 238  & 11.42(0.21) & 0.75(0.04) & 9.45(0.19)  & 10.82(0.27) & 3.27 & 45.23(30.76,5.62)     & \(<0.25\)	  \\ 
24 & 59   & 11.45(0.23) & 0.83(0.01) & 9.39(0.15)  & 10.83(0.24) & 3.78 & 111.59(38.68,0.45)    & -0.33(0.17) \\
25 & 319  & 11.47(0.21) & 0.86(0.02) & 9.40(0.16)  & 10.84(0.28) & 3.04 & 65.09(47.91,3.10)     & \(>0.11\)   \\
26 & 122  & 11.52(0.21) & 0.93(0.01) & 9.37(0.15)  & 10.84(0.30) & 3.25 & 104.73(50.43,2.76)    & -0.33(0.50) \\
27 & 565  & 11.48(0.20) & 1.06(0.08) & 9.33(0.12)  & 10.66(0.49) & 3.04 & 204.40(159.37,38.54)  & \(<0.14\)	  \\
28 & 24   & 11.62(0.18) & 1.20(0.01) & 9.40(0.06)  & 10.31(0.82) & 3.26 & 1234.10(428.64,4.11)  & \(>0.57\)	  \\
29 & 3    & 11.48(0.12) & 1.21(0.01) & 9.40(0.03)  & 10.85(0.14) & 3.09 & 606.33(204.05,0.18)   & 0.70(0.27)  \\
30 & 112  & 11.50(0.18) & 1.25(0.04) & 9.33(0.13)  & 10.50(0.69) & 3.02 & 653.77(277.84,46.88)  & \(>0.31\)	  \\
31 & 80   & 11.55(0.18) & 1.40(0.04) & 9.30(0.13)  & 10.56(0.64) & 1.29 & \(<1299.87\)          & \(>-0.89\)  \\
32 & 8    & 12.05(0.03) & 0.66(0.09) & 9.75(0.12)  & 11.59(0.05) & 6.55 & 368.43(62.19,123.45)  & -0.18(0.16) \\
33 & 27   & 12.10(0.09) & 1.09(0.18) & 9.41(0.09)  & 11.41(0.11) & 2.31 & 332.98(211.58,137.06) & -0.30(0.67) \\
\hline
\hline
\end{tabular}
}
\end{center}
\end{table}

\begin{table}
\begin{center}
\caption{Same as Table \ref{tab:stack} but for redshift selected stacking bins.}\label{tab:stack_z}
\resizebox{\textwidth}{!}{
\begin{tabular}{ccccccccc}
\hline
\hline
(1) bin number & (2) number of sources & (3) \(<{\log({L_K/L_{\astrosun}})}>\) & (4) \(<{z}>\) & (5) \(<{\log({\mbox{age}/\mbox{yr}})}>\) & (6) \(<{\log({M_*/M_{\astrosun}})}>\) & (7) \(S/N\) & (8) \(<L_{X-LMXB,\,0.3-8}/({10}^{40}\mbox{ erg}\mbox{ s}^{-1})>\) & (9) HR  \\
\hline
1  & 186  &  9.12(0.07) & 0.29(0.10) & 9.53(0.31) &  8.66(0.25) & 3.02 & 0.96(0.36,0.72)       & 0.54(0.33)  \\
2  & 511  &  9.84(0.23) & 0.31(0.10) & 9.52(0.26) &  9.28(0.34) & 3.03 & 0.68(0.28,0.46)       & -0.33(0.52) \\
3  & 129  & 10.42(0.06) & 0.34(0.10) & 9.49(0.24) &  9.96(0.12) & 3.07 & 1.88(0.85,1.20)       & \(>0.17\)   \\
4  & 241  & 10.75(0.10) & 0.35(0.08) & 9.56(0.22) & 10.29(0.15) & 3.01 & 1.08(0.86,0.57)       & 0.12(0.58)  \\
5  & 97   & 11.01(0.04) & 0.35(0.08) & 9.61(0.23) & 10.57(0.09) & 3.11 & 1.77(1.33,0.94)       & \(>-0.34\)  \\
6  & 9    & 11.09(0.01) & 0.32(0.08) & 9.76(0.23) & 10.68(0.09) & 3.38 & 6.38(2.36,3.77)       & -0.04(0.36) \\
7  & 91   & 11.17(0.04) & 0.36(0.07) & 9.70(0.22) & 10.74(0.09) & 3.06 & 2.38(1.84,1.03)       & \(<-0.12\)  \\
8  & 37   & 11.27(0.02) & 0.36(0.07) & 9.74(0.20) & 10.85(0.08) & 3.06 & 4.25(2.42,2.08)       & 0.07(0.48)  \\
9  & 46   & 11.35(0.03) & 0.35(0.07) & 9.81(0.19) & 10.95(0.08) & 3.03 & 2.85(2.06,1.30)       & \(<-0.10\)  \\
10 & 25   & 11.43(0.02) & 0.37(0.07) & 9.80(0.20) & 11.02(0.08) & 3.02 & 6.17(3.61,2.70)       & \(>0.42\)   \\
11 & 20   & 11.48(0.01) & 0.38(0.07) & 9.82(0.17) & 11.09(0.07) & 3.10 & 8.66(4.38,3.50)       & -0.37(0.50) \\
12 & 18   & 11.52(0.02) & 0.36(0.07) & 9.84(0.13) & 11.10(0.06) & 3.25 & 5.70(3.02,2.77)       & -0.07(0.42) \\
13 & 28   & 11.59(0.03) & 0.36(0.07) & 9.86(0.10) & 11.20(0.05) & 3.05 & 3.96(2.80,1.82)       & \(<0.50\)   \\
14 & 13   & 11.69(0.02) & 0.37(0.06) & 9.85(0.16) & 11.27(0.10) & 3.13 & 8.90(5.04,3.12)       & \(<-0.33\)  \\
15 & 18   & 11.81(0.05) & 0.38(0.06) & 9.83(0.07) & 11.37(0.09) & 3.05 & 6.00(4.60,2.10)       & -0.01(0.53) \\
16 & 425  &  9.83(0.24) & 0.73(0.14) & 9.38(0.37) &  9.09(0.46) & 3.02 & 43.08(15.98,20.05)    & 0.52(0.36)  \\
17 & 1159 & 10.91(0.18) & 0.77(0.13) & 9.33(0.17) & 10.26(0.38) & 3.07 & 21.41(15.95,9.18)     & 0.05(0.54)  \\
18 & 262  & 11.28(0.04) & 0.79(0.13) & 9.41(0.20) & 10.74(0.15) & 3.03 & 58.62(32.17,23.11)    & -0.33(0.42) \\
19 & 34   & 11.36(0.01) & 0.82(0.14) & 9.45(0.21) & 10.79(0.21) & 3.30 & 131.37(47.08,52.96)   & 0.11(0.33)  \\
20 & 111  & 11.40(0.02) & 0.81(0.14) & 9.44(0.21) & 10.86(0.11) & 3.04 & 103.10(48.81,43.24)   & 0.20(0.42)  \\
21 & 106  & 11.47(0.02) & 0.81(0.14) & 9.45(0.21) & 10.93(0.14) & 3.08 & 115.21(55.10,47.07)   & 0.50(0.47)  \\
22 & 274  & 11.61(0.05) & 0.80(0.14) & 9.53(0.22) & 11.08(0.11) & 3.07 & 52.12(37.39,21.70)    & \(>-0.29\)  \\
23 & 8    & 11.72(0.01) & 0.77(0.10) & 9.53(0.15) & 11.22(0.12) & 3.17 & 204.51(79.18,65.47)   & 0.18(0.37)  \\
24 & 51   & 11.75(0.02) & 0.79(0.15) & 9.60(0.24) & 11.24(0.10) & 3.78 & 102.39(36.38,46.82)   & \(<-0.59\)  \\
25 & 86   & 11.84(0.04) & 0.81(0.14) & 9.59(0.20) & 11.32(0.10) & 3.03 & 118.48(66.52,48.44)   & \(<0.09\)   \\
26 & 28   & 11.95(0.02) & 0.85(0.12) & 9.57(0.19) & 11.41(0.09) & 3.24 & 149.06(60.70,33.54)   & 0.04(0.39)  \\
27 & 15   & 12.04(0.02) & 0.81(0.16) & 9.60(0.21) & 11.50(0.10) & 6.03 & 397.82(77.91,190.80)  & -0.30(0.20) \\
28 & 8    & 12.13(0.03) & 0.86(0.16) & 9.52(0.17) & 11.50(0.09) & 1.20 & \(<269.42\)           & \(-0.16*\)  \\
29 & 737  & 11.14(0.27) & 1.18(0.12) & 9.30(0.20) & 9.93(0.79)  & 3.06 & 402.00(188.34,102.88) & \(>0.20\)   \\
30 & 127  & 11.68(0.06) & 1.20(0.12) & 9.36(0.10) & 10.90(0.44) & 3.25 & 404.13(176.27,100.36) & -0.23(0.44) \\
31 & 14   & 11.84(0.02) & 1.25(0.15) & 9.34(0.11) & 11.19(0.16) & 3.18 & 299.63(107.66,90.89)  & -0.46(0.41) \\
32 & 33   & 12.01(0.13) & 1.21(0.13) & 9.38(0.05) & 11.24(0.41) & 2.32 & 557.18(342.18,144.63) & \(<0.36\)   \\
\hline
\hline
\end{tabular}
}
\end{center}
\end{table}

\end{document}